\documentclass[%
 ,twocolumn%
 ,secnumarabic%
,superscriptaddress,amssymb, amsmath,nobibnotes, aps, prc, showpacs]{revtex4}
\usepackage{bm}%
\usepackage{graphicx}

\expandafter\ifx\csname package@font\endcsname\relax\else
 \expandafter\expandafter
 \expandafter\usepackage
 \expandafter\expandafter
 \expandafter{\csname package@font\endcsname}%
\fi

\begin{document}

\title{Paris $N\bar N$ potential constrained by recent antiprotonic-atom data\\
and $\bar np$ total cross sections}%

\author{B. El-Bennich}
\affiliation{Laboratoire de Physique Nucl\'eaire et de Hautes \'Energies, Groupe Th\'eorie, IN2P3-CNRS, Universit\'es Pierre \& Marie Curie et Paris Diderot, 4 Place Jussieu, 75252 Paris, Cedex, France }
\affiliation{Physics Division, Argonne National Laboratory, Argonne, IL 60439, USA}
\author{M. Lacombe}
\affiliation{Laboratoire de Physique Nucl\'eaire et de Hautes \'Energies, Groupe Th\'eorie, IN2P3-CNRS, Universit\'es Pierre \& Marie Curie et Paris Diderot, 4 Place Jussieu, 75252 Paris, Cedex, France }
\author{B. Loiseau}
\affiliation{Laboratoire de Physique Nucl\'eaire et de Hautes \'Energies, Groupe Th\'eorie, IN2P3-CNRS, Universit\'es Pierre \& Marie Curie et Paris Diderot, 4 Place Jussieu, 75252 Paris, Cedex, France }
\author{S. Wycech}
\affiliation{Soltan Institute for Nuclear Studies, Warsaw, Poland}

\begin{abstract}
We report on an updated Paris $N\bar N$  optical potential.
The long- and intermediate-range real parts are obtained by $G$-parity transformation of the Paris $NN$ potential based on a theoretical dispersion-relation treatment of the correlated and uncorrelated two-pion exchange.
The short-range imaginary potential parametrization results from the calculation of the $N\bar N $ annihilation box diagram into two mesons with a nucleon-antinucleon intermediate state in the crossed channel.
The parametrized real and imaginary short range parts are determined by fitting not only the existing experimental data included in the 1999 version of the Paris $N\bar N$ potential, but also the recent antiprotonic-hydrogen data and $\bar np$ total cross sections.
The description of these new observables is improved.
Only this readjusted potential generates an isospin zero $^1S_0$, 52 MeV broad quasibound state at 4.8 MeV below the threshold.
Recent BES data on $J/\psi$ decays could support the existence of such a state.
\end{abstract}
\maketitle

\section{Introduction \label{introduction}}
There has been recently a renewal of interest in the nucleon-antinucleon, $N\bar N$, interaction due to the observation of near threshold enhancements in the proton-antiproton, $p\bar p$, invariant mass spectrum of heavy meson decays such as $J/\psi\to\gamma p\bar p$~\cite{Bai2003}, $B^+\to p\bar p K^+,\ B^+\to p\bar p\pi^+$~\cite{Wei2007}, $B^0\to p\bar pK^{*0}$~\cite{Chen2008} and $\bar B^0\to D^{(*)0}p\bar p$~\cite{Abe2002}.
On the other hand, no such structure was observed by the BES Collaboration for the $J/\psi\to\pi^0p\bar p$ decays~\cite{Bai2003}.
For the radiative and pionic $J/\psi$ decays reported in Ref.~\cite{Bai2003}, two of us have proposed a natural explanation following from a traditional model of $p\bar p$ interactions~\cite{Loiseau2005}.
These interactions originate from the Paris  $N\bar N$ potentials~\cite{Paris82,Paris94,Paris99}.
Taking into account the low energy allowed final states, the BES data are well reproduced with an isospin one $p\bar p(^1P_1)$ wave for the $\pi^0p\bar p$ channel and a $p\bar p(^1S_0)$ wave for the $\gamma p\bar p$ channel.
It was furthermore shown in Ref.~\cite{Loiseau2005} that the best results were obtained with an upgraded $N\bar N$ Paris potential constrained not only by the set of data used in the 1999 version~\cite{Paris99}, but also by recent total $\bar np$ cross sections of Ref.~\cite{Iazzi2000} and antiproton-hydrogen widths and shifts~\cite{Augsburger99,Gotta99}.
Only this recently readjusted potential has an isospin $T=0$ $^1S_0$ quasibound state close to the $p\bar p$ threshold.
The existence of such a state has some support from the BES data even if the low-energy $p\bar p$ spectrum of the radiative decay could also be reproduced in Ref.~\cite{Sibirtsev2005} using the $T=1$ $S$-wave of the meson-exchange J\"ulich-Bonn $N \bar N$ model where no $^1S_0$ bound state is present.
The aim of the present work is to report on the updated Paris $N\bar N$ potential used in Ref.~\cite{Loiseau2005}.

\begin{table}
\caption{Heights of the different real potentials $U(r)$ at $r=r_3=0.188$ fm and $r=r_2=0.587$ fm
together with the parameters $g$ (dimensionless) and $f$ of the imaginary potentials. 
These quantities, determined by
the fit to experimental observables, are compared with those of the Paris
99 potential~\cite{Paris99}.
All $U(r)$ are in MeV but the $U^b(r)$ which are dimensionless.
The definitions of
the real and imaginary potentials can be found in the Appendix. \label{potparam}}
\begin{ruledtabular}
\begin{tabular}{l|cc|cc}
                 & \multicolumn{2}{c|}{Isospin $T=0$} & \multicolumn{2}{c}{Isospin $T=1$} \\
           & This work & Paris 99 & This work & Paris 99 \\
\hline
$U_0^a(r_3)$ & 8692.49 & 8594.41 & -5300.64 &  -1917.54  \\
$U_0^a(r_2)$ & -378.44  & -489.08 & -664.40 &  -1716.76 \\
$U_0^b(r_2)$ &  0.5857 & 1.307  & -0.327 &  -0.132 \\
$U_1^a(r_3)$ &  -6508.93 & -5286.67 & 5001.23 & 3121.01 \\
$U_1^a(r_2)$ &  -1041.26  & -810.89  & -1115.79  & -1135.07 \\
$U_1^b(r_2)$ &-1.306 & -1.741 & -1.676 & -1.931 \\
$U_{LS}(r_2)$ & 917.12 & 788.30 & -436.31 & -423.71 \\
$U_{T}(r_2)$ & 481.68 & 397.14 & 216.46 & 128.14 \\
$U_{SO2}(r_2)$ & 105.43 & 75.03 &  203.28 & 172.48 \\
$g_c$ & 153.57 & 124.86 & 153.82 & 78.40 \\
$f_c$(MeV$^{-1}$) & 0.0153  & 0.0190 & 0.0121 & 0.0335 \\
$g_{SS}$ & -15.56 & -3.83 & 45.49 & 19.94 \\
$f_{SS}$(MeV$^{-1}$) & 0.0076 & -0.0373  & 0.0135 & 0.0412 \\
$g_{LS}$ & 0.010 & 35.369 & 0.026 & 12.027 \\
$g_T$ & 0.023 & 2.057 & 0.027 & 5.073 \\
\end{tabular}
\end{ruledtabular}
\end{table}

\begin{figure*}[ht]
\includegraphics[scale=0.45]{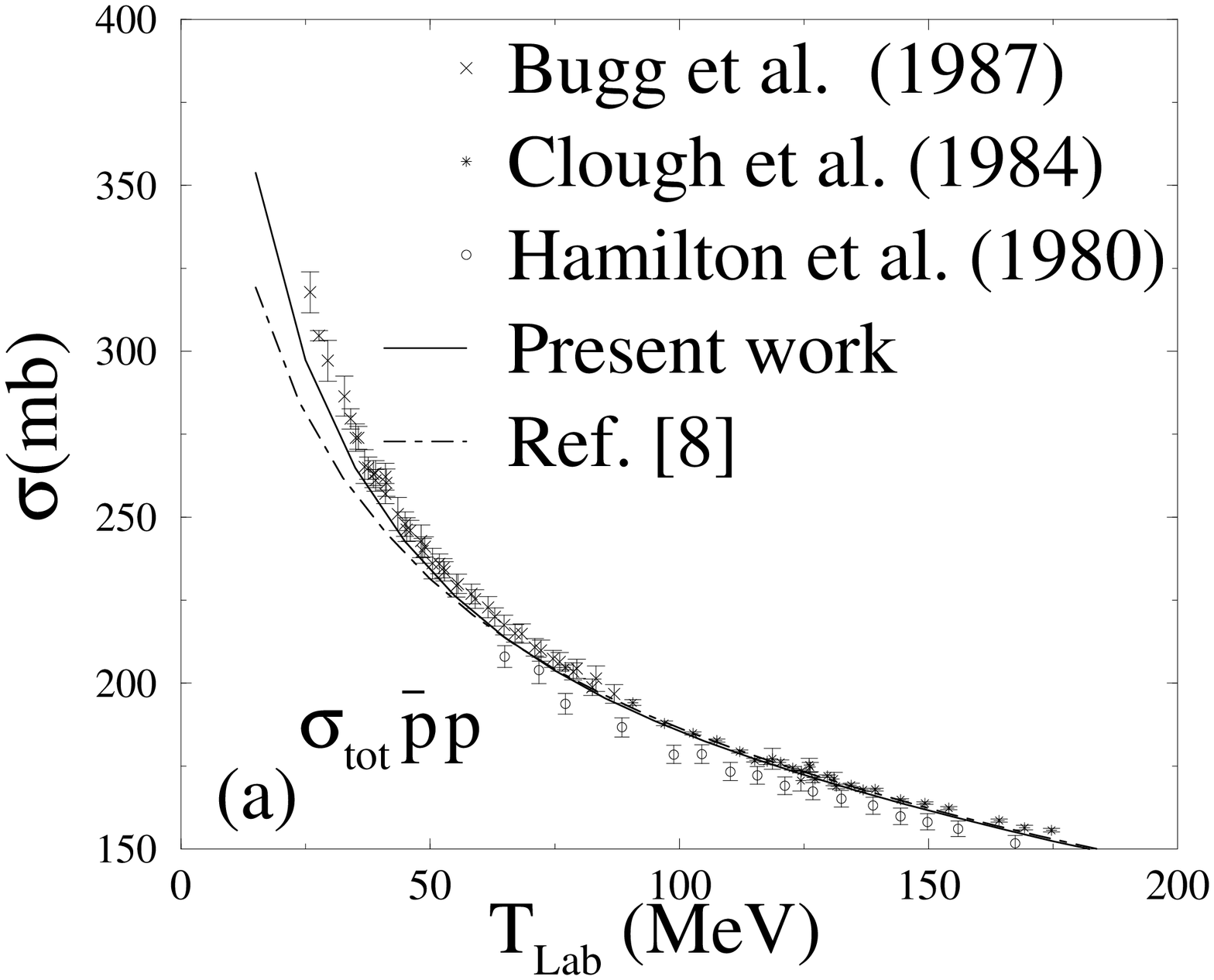}
\includegraphics[scale=0.45]{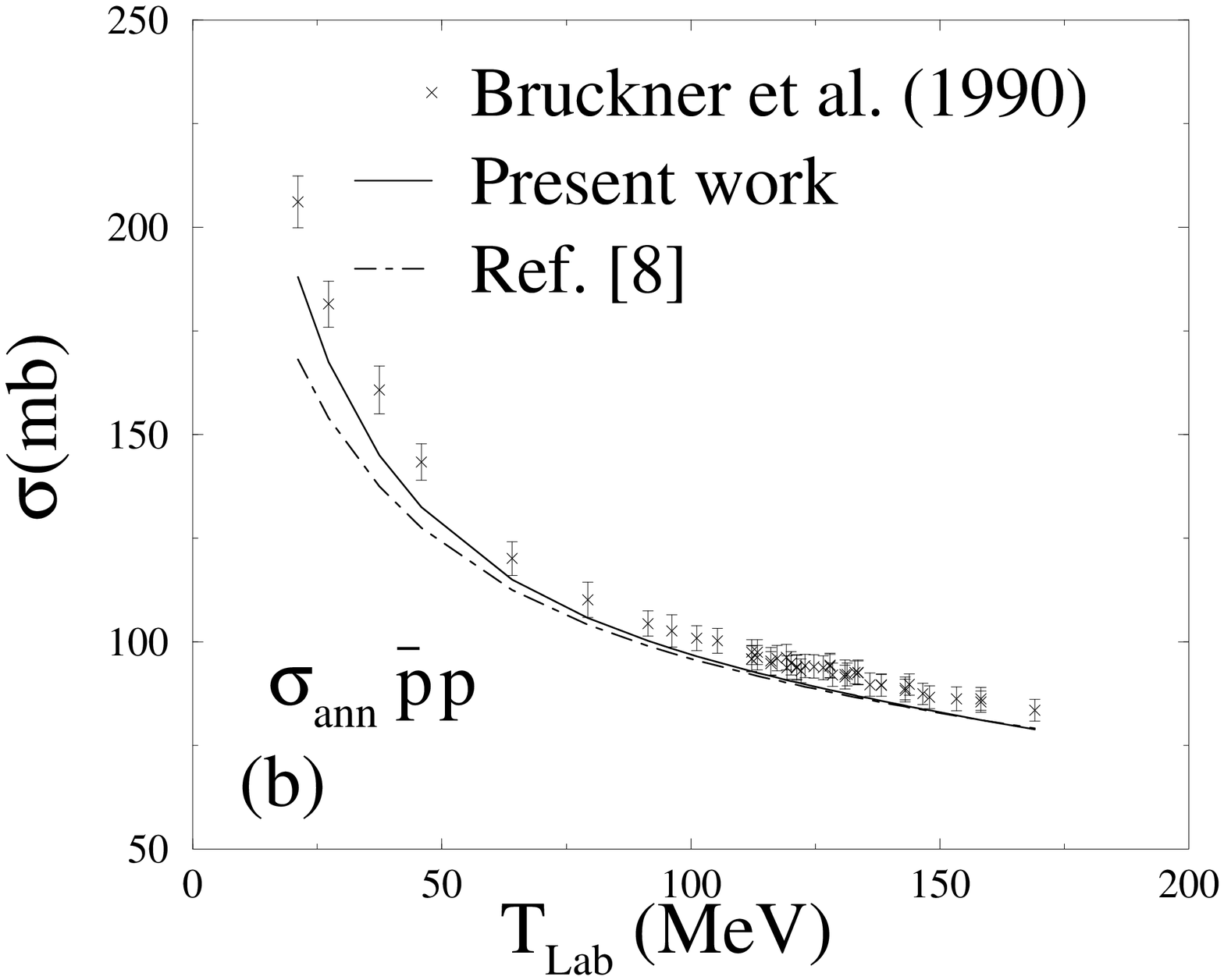}
\includegraphics[scale=0.44]{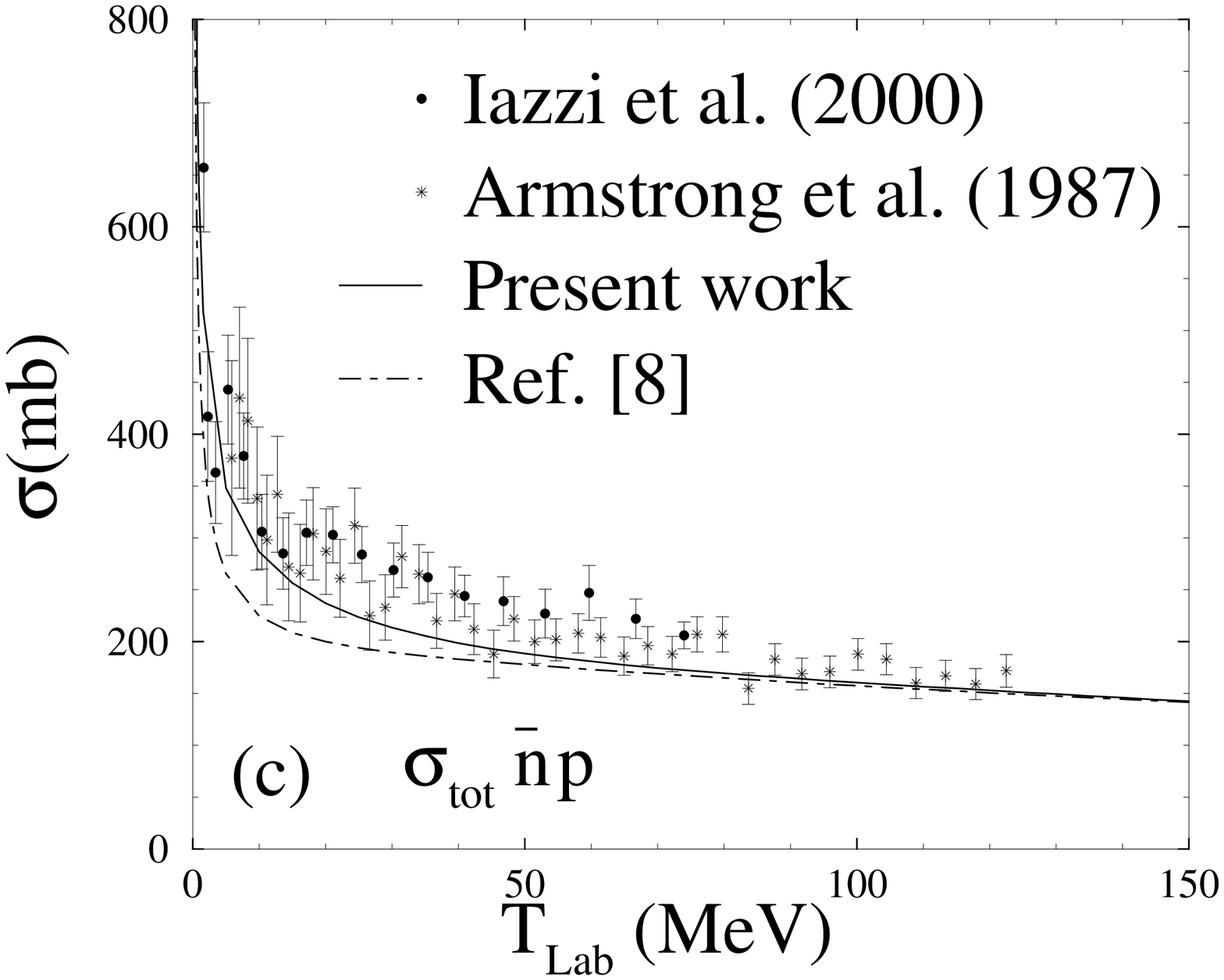}
\includegraphics[scale=0.44]{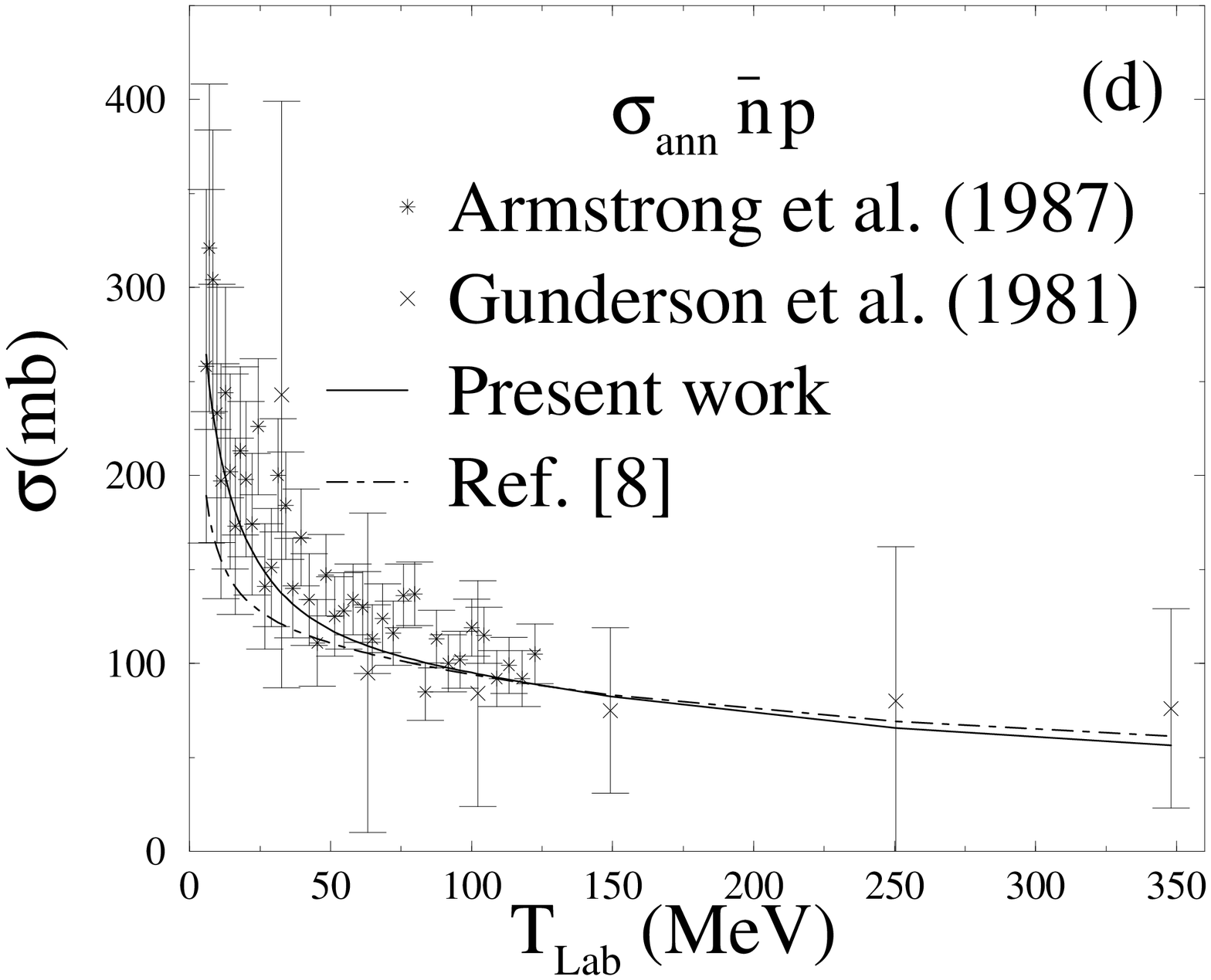}
\caption{\label{Fig1}Total and annihilation cross sections for the $\bar pp$ and $\bar np$ systems. The references of the experimental data can be found in Ref.~\cite{Paris94}. The data of Iazzi \textsl{et al.} in (c) are from Ref.~\cite{Iazzi2000}.}
\end{figure*}

The paper is organized as follows. 
In Sec.~\ref{modelandresults}, after a brief reminder of the model, we compare its results to the experimental scattering observables and to the results of the 1999 $N \bar N$ Paris potential.
This comparison is also done for the antiprotonic-hydrogen level shifts and widths.
The close to threshold bound states and resonances of the present work are searched for and compared to those of the 1999 model.
We then plot and compare the optical potentials of both solutions.
Section~\ref{discussion} is devoted to a discussion of our results.
A summary and conclusions are presented in Sec.~\ref{summary}. 
Finally, the Appendix reminds the reader of the full expression of the optical potential.

\begin{figure*}[ht]
\includegraphics[angle=-90,scale=0.35]{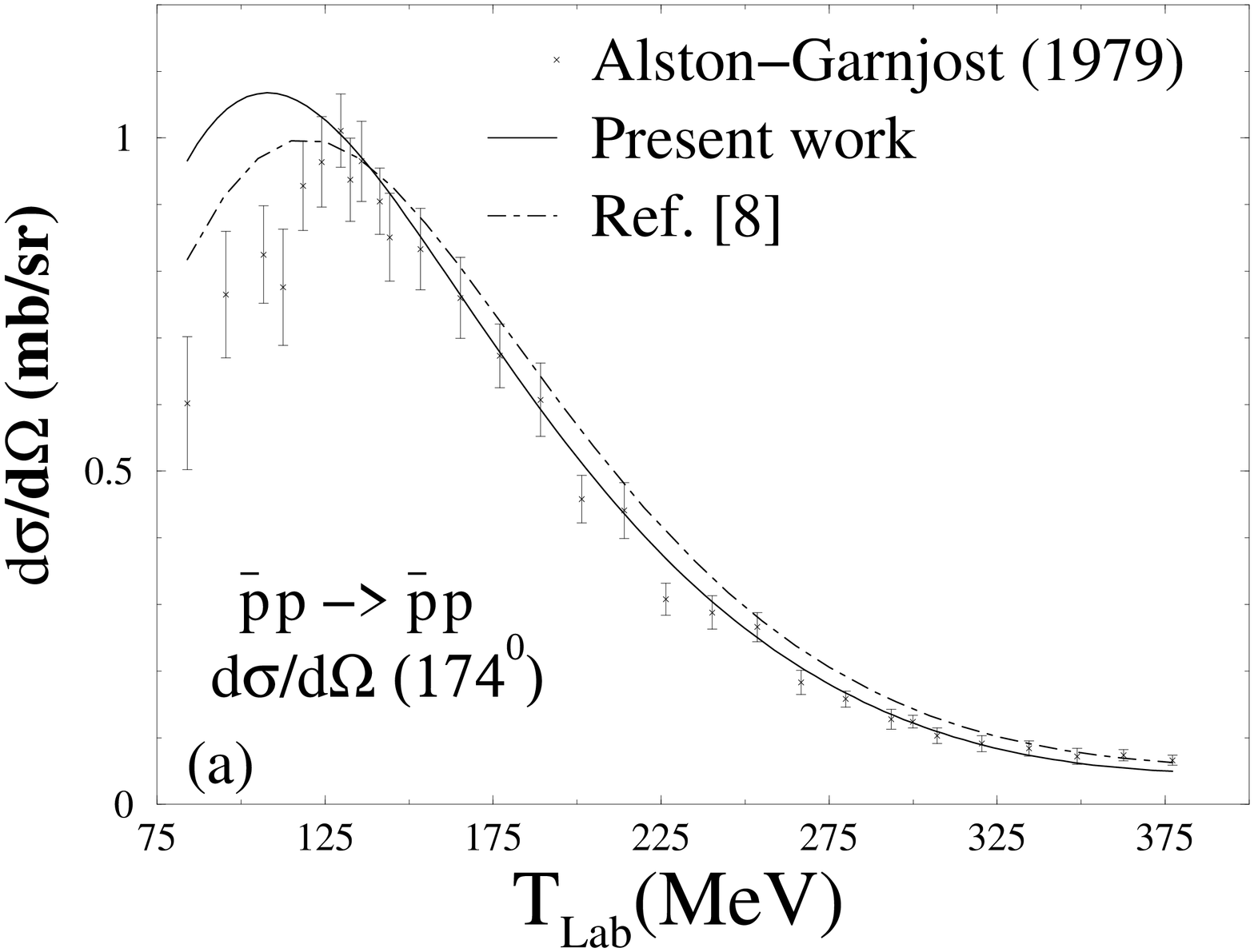}
\includegraphics[angle=-90,scale=0.35]{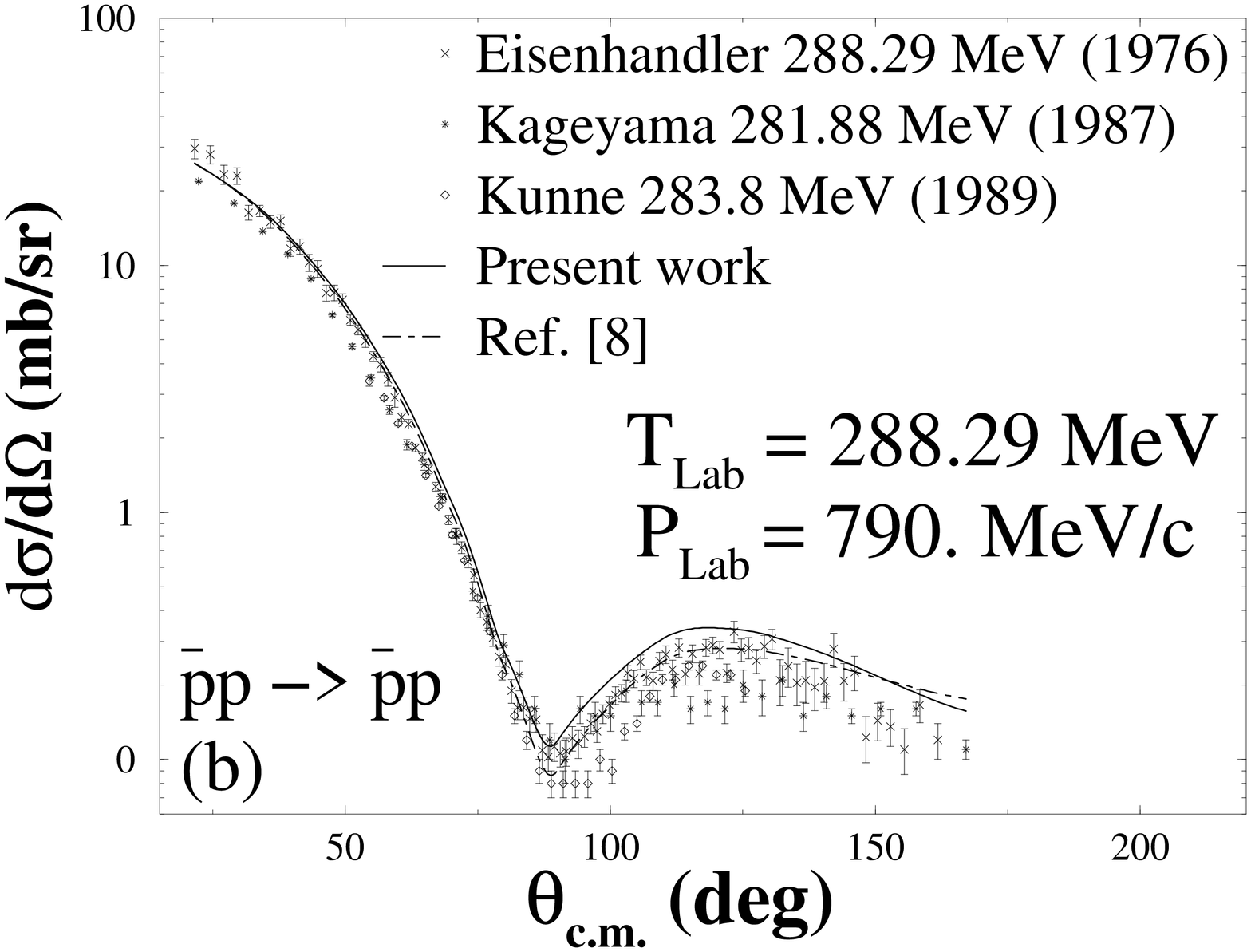}
\includegraphics[angle=-90,scale=0.35]{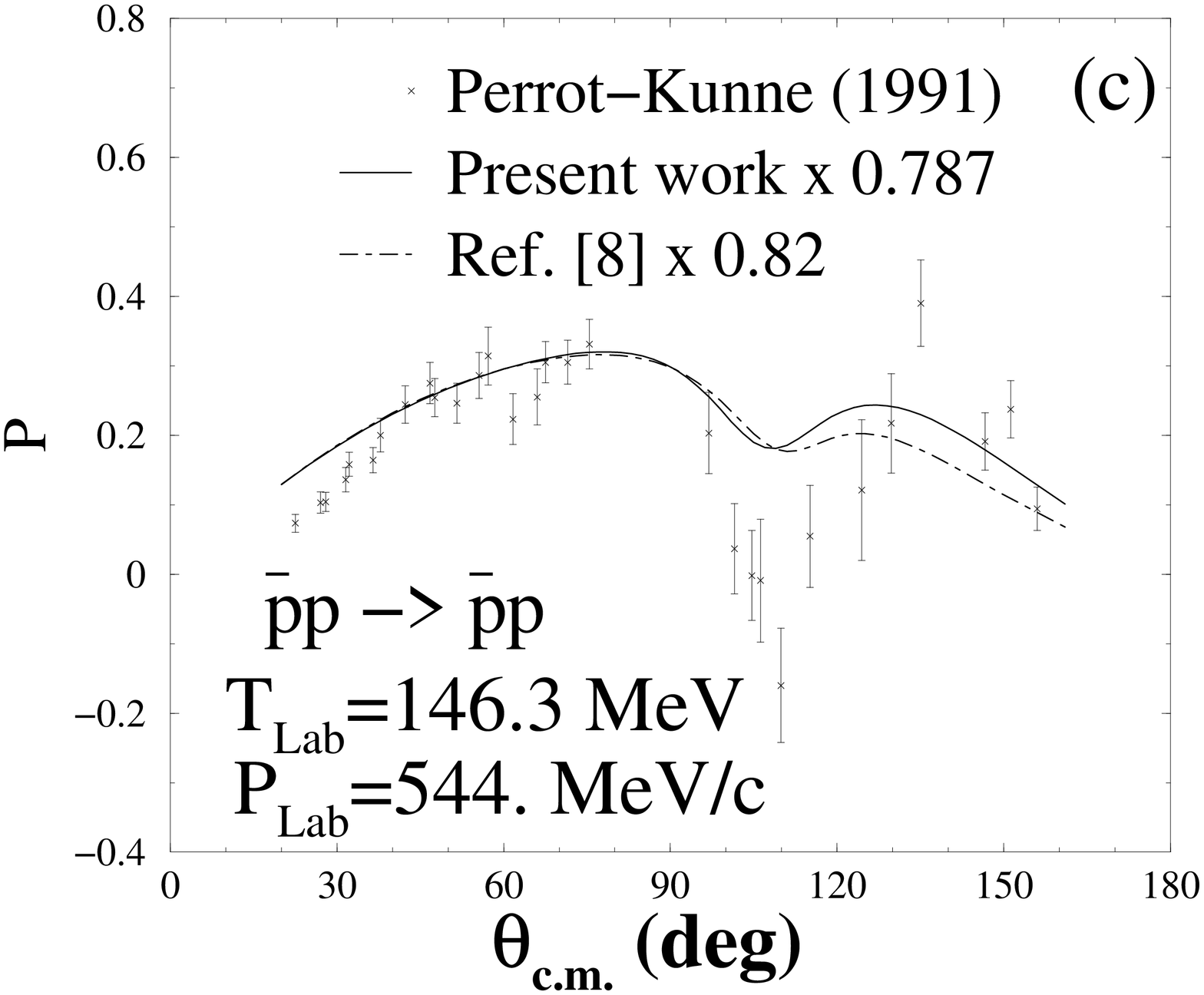}
\includegraphics[angle=-90,scale=0.35]{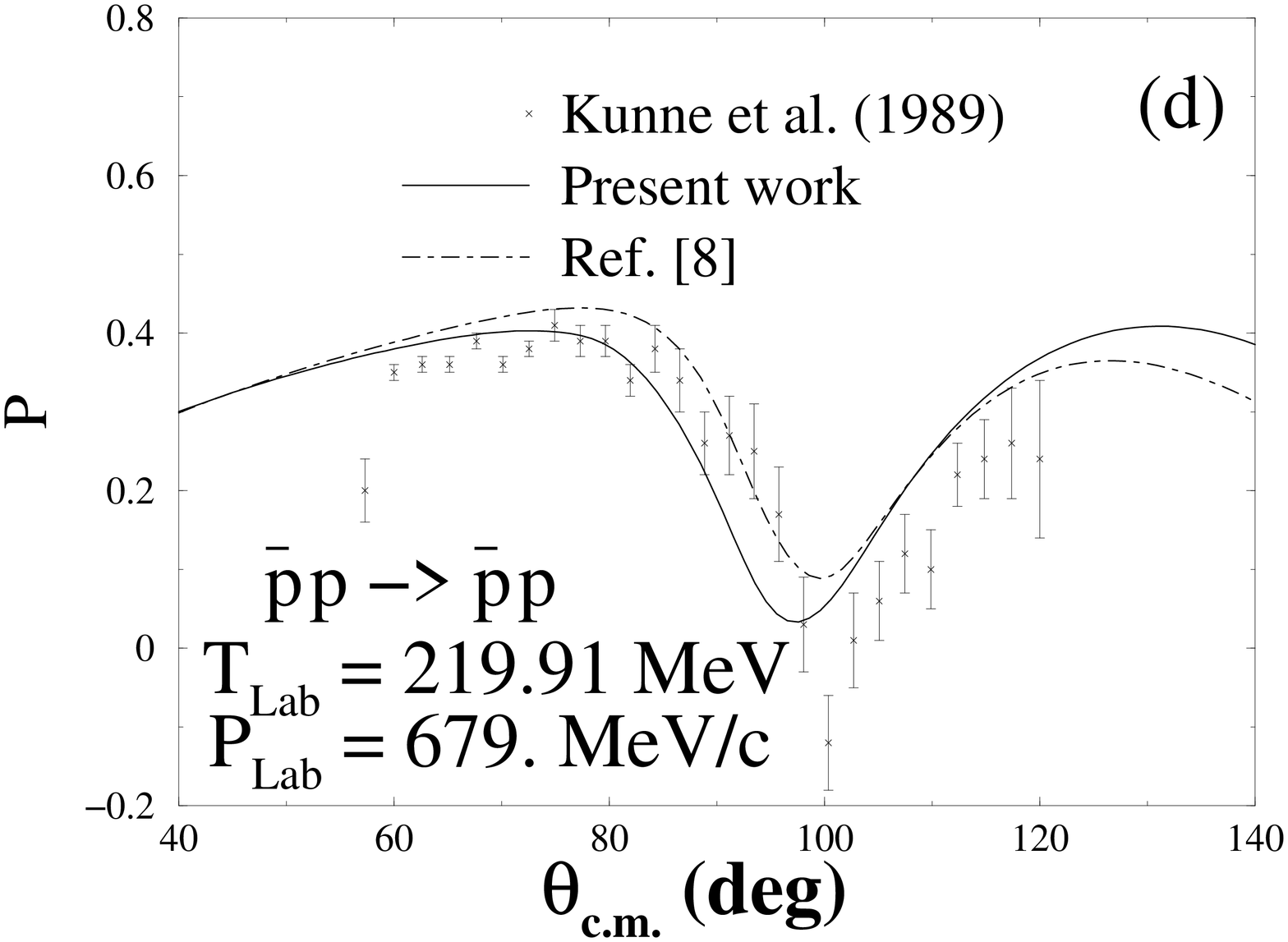}
\caption{\label{alston} Differential cross sections and polarization for the $\bar pp$  systems. The references to the experimental data can be found in Ref.~\cite{Paris94}.}
\end{figure*}

\section{Model and Results \label{modelandresults}}

\subsection{Brief reminder of the model \label{model}}

The 1982 Paris $N\bar N$ optical potential~\cite{Paris82} was itself readjusted in 1994~\cite{Paris94} and 1999~\cite{Paris99}.
For all these potentials and the present one, the $N \bar N$ interaction is described by an energy dependent optical potential
\begin{equation}
\label{opticalpotential}
V_{N \bar N} \left({\bf r}, T_{Lab}\right)=
U_{N \bar N}\left({\bf r}, T_{Lab}\right) 
-i\ W_{N \bar N} \left({\bf r}, T_{Lab}\right),
\end{equation}
where the nonlocality of the real $U_{N \bar N}$ and imaginary $W_{N \bar N}$ potentials are accounted for by a linear energy dependence in the kinetic energy $T_{Lab}$.
Meson exchanges explain in a satisfactory way the $NN$ force for large and medium distances between the nucleons.
Therefore the long and intermediate range real parts, i.e. those for inter $N\bar N$ distances $r \geq 1$ fm, are obtained by the $G$-parity transformation of the corresponding parts of the Paris $NN$ potential~\cite{Paris80} based on a theoretical dispersion-relation  treatment of the correlated and uncorrelated two-pion exchange~\cite{Cottingham73}.
These real potentials contain, besides the one-pion exchange, the two-pion exchange and the $\omega$ and $A_1$ meson exchanges as parts of the three-pion exchange.
For $ r < 1$ fm heavier meson exchanges and/or other degrees of freedom, such as
quarks and gluons take place but the available theoretical calculations are not free from phenomenological parameters~(see for instance Refs.~\cite{Klempt2002} and \cite{Entem2006}).
Consequently, following the choice made in the case of the Paris $NN$ potential~\cite{Paris80}, we use here an empirical short range real potential.

As in Ref.~\cite{Paris94}, we expand for $r<1$ fm the phenomenological radial potentials in power of 
$r$ [Eqs.~(\ref{cubic}) and (\ref{quadratic})] and match them to the theoretical ones at two point in the vicinity of 1 fm. 
Then, above 1 fm the theoretical potentials are entirely preserved.
For each isospin state, the spin structure of the $N \bar N$ interaction requires five independent invariants with five radial potentials.
A phenomenological cubic expansion is used for the central components [see Eq.~(\ref{cubic})] and a quadratic one [see Eq.~(\ref{quadratic})] for the other terms.
This  leads to nine parameters representing the strength of the different empirical potentials at two (for the central components) or one (for the other) specific $r$ values smaller than 1~fm (see Table~\ref{potparam}).
The expressions of the full real potentials and more details on our fitting procedure are given in the Appendix.

\begin{figure*}[ht]
\includegraphics[angle=-90,scale=0.34]{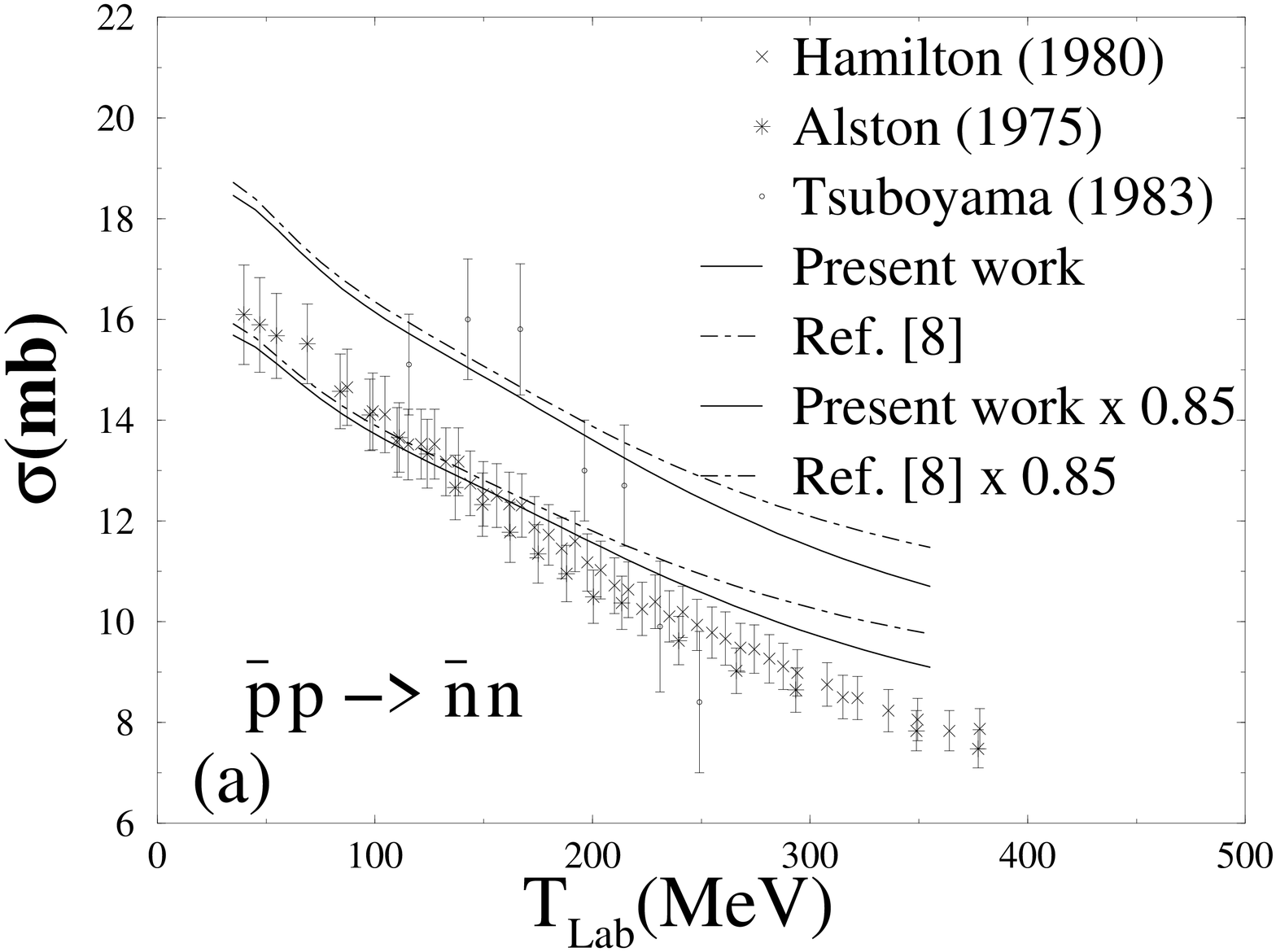}
\includegraphics[angle=-90,scale=0.35]{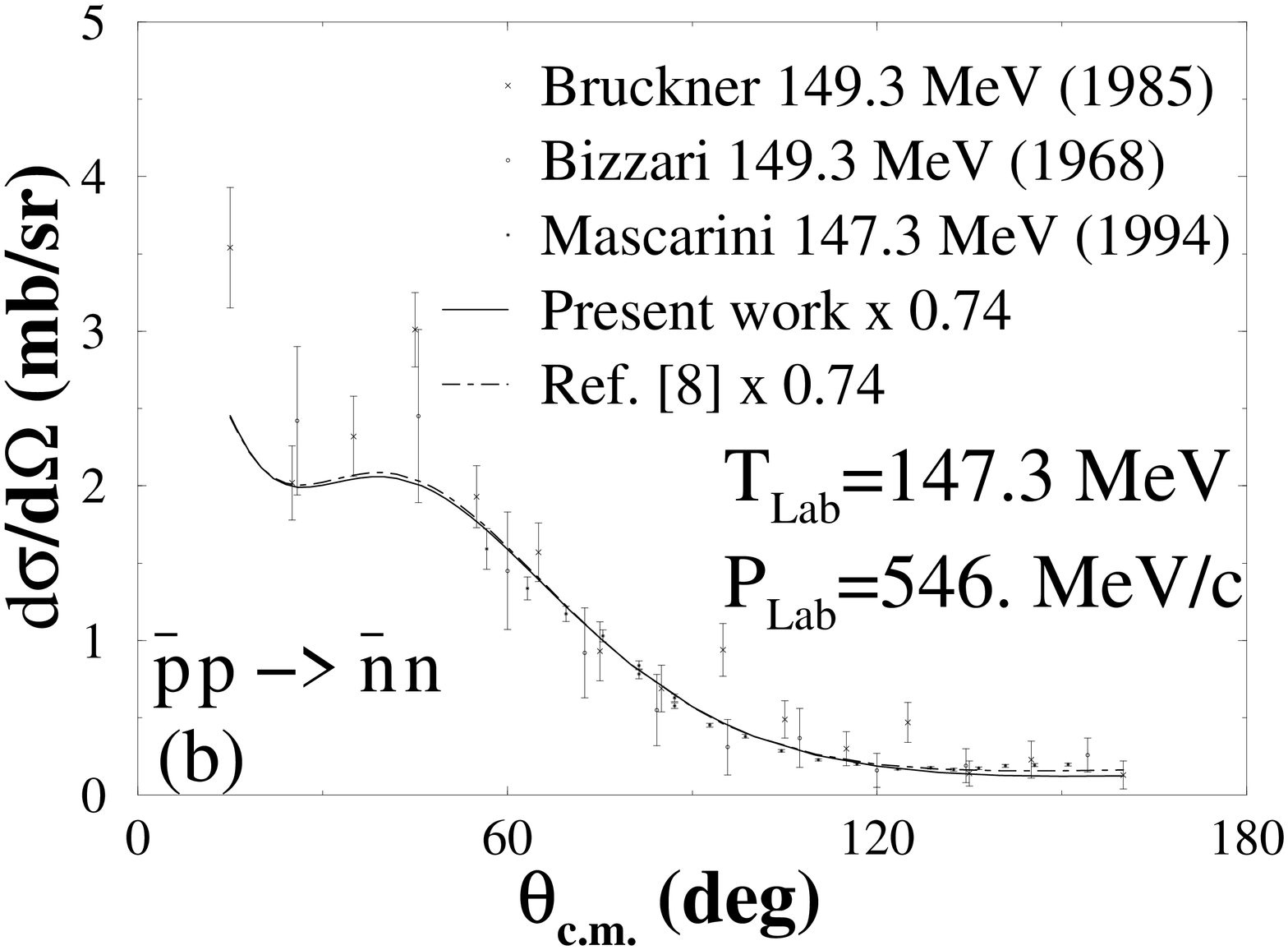}
\includegraphics[angle=-90,scale=0.345]{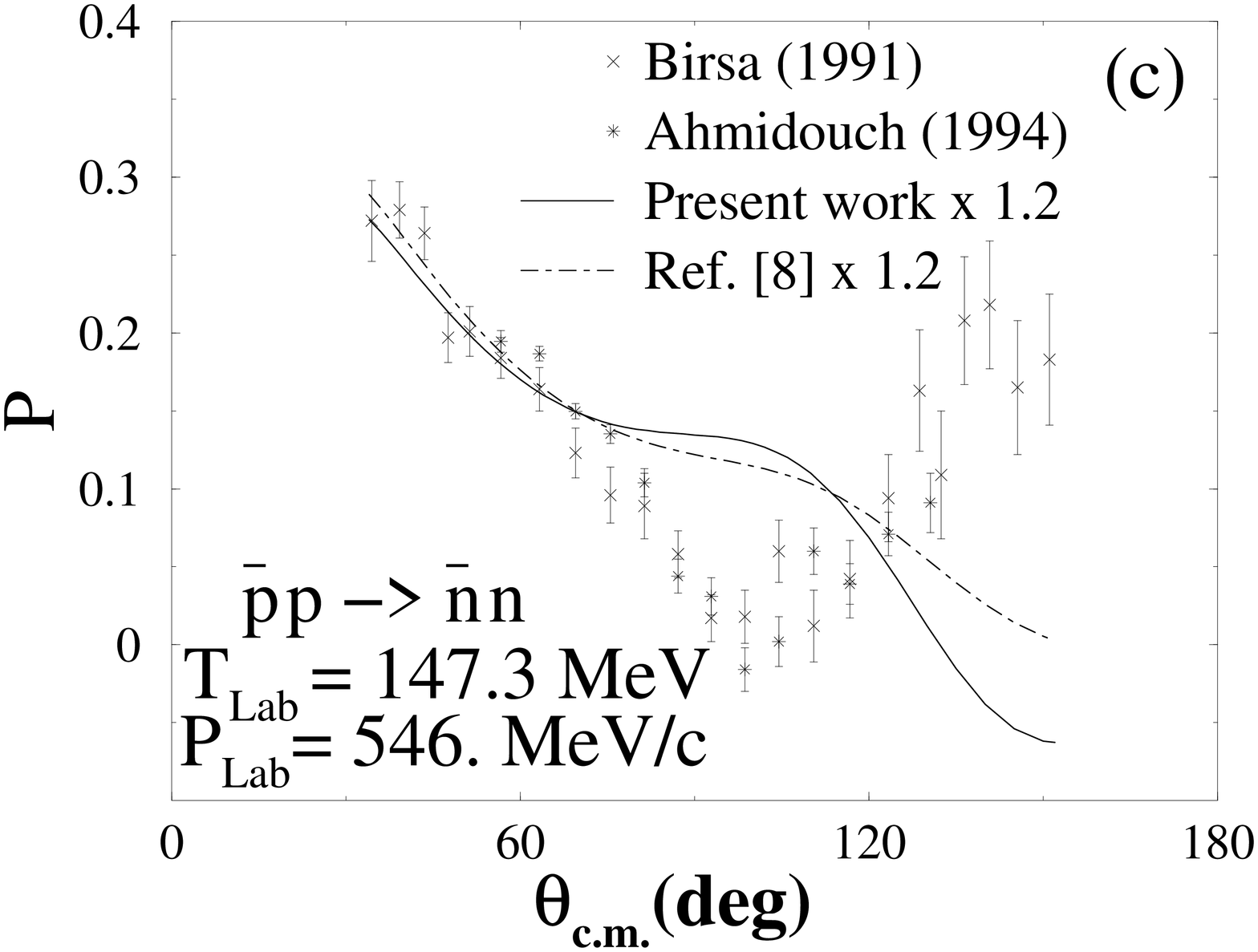}
\includegraphics[angle=-90,scale=0.345]{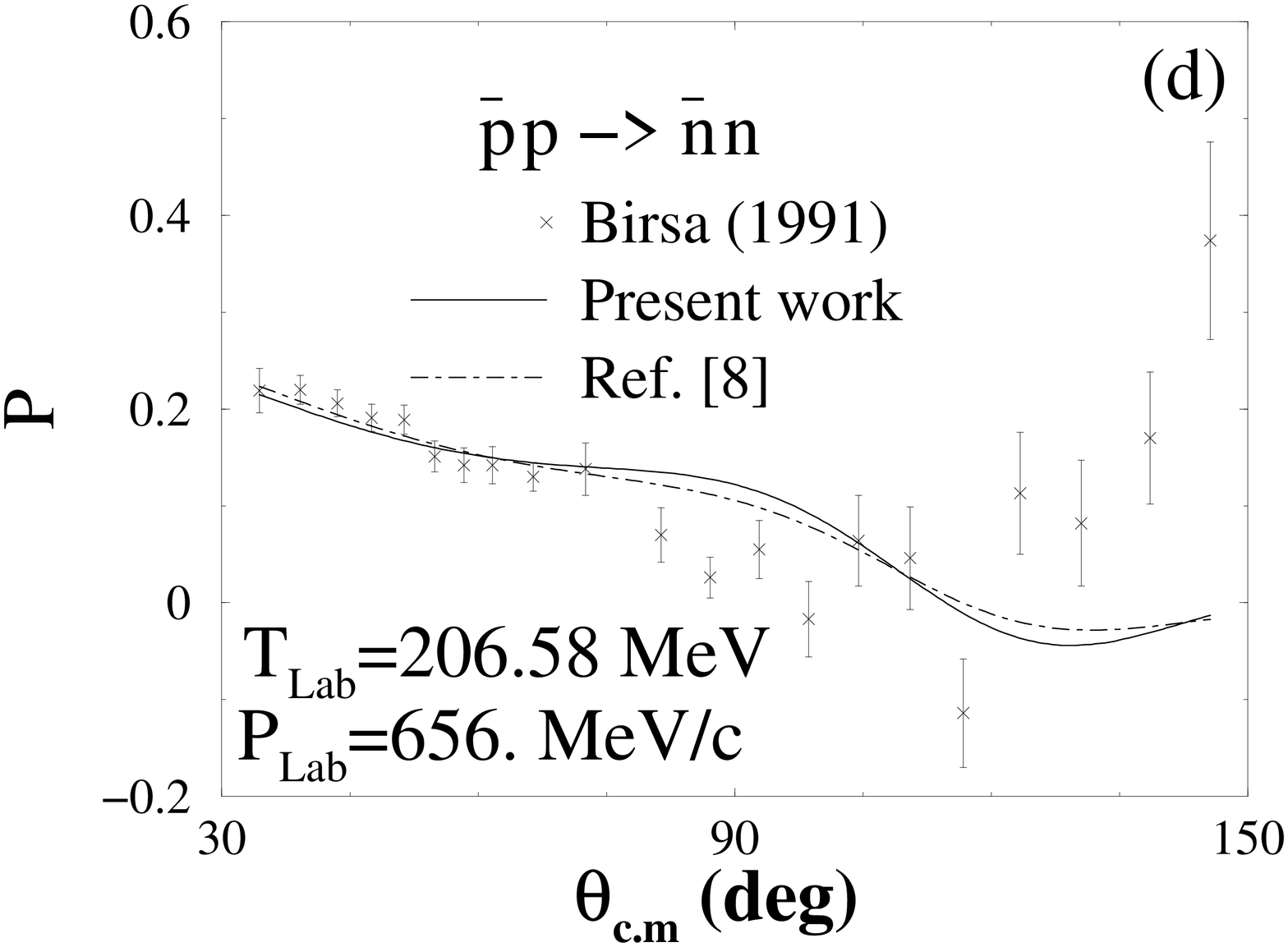}
\caption{\label{cex} Total and differential cross sections and polarization for the $\bar pp \to \bar nn$ systems. The references to the experimental data can be found in Ref.~\cite{Paris94}.}
\end{figure*}

The imaginary potential $W_{N\bar N}$ includes the $N \bar N$ annihilation into mesons in the $s$-channel and can be calculated from annihilation diagrams~\cite{Paris94,Klempt2002}.
It implies the exchange of a baryon-antibaryon pair in the crossed $t$-channel and the resulting potential is then non-local and short range.
We give in the Appendix the full expression we used.
More details of its derivation can be found in the Appendix of Ref.~\cite{Paris94}.

The parameters of the real and imaginary short range parts are then determined by fitting the existing experimental data, viz. 915 data points in 1982, 3800 in 1994 and over 4000 in 1999.
References and discussions on the experimental 1999 data set is given in Ref.~\cite{Paris99}.
A recent review on the $N \bar N$ data and the underlying physics can also be found in Ref.~\cite{Klempt2002}.
Our data consist of  4259 data used in the model 1999~\cite{Paris99} plus the 64 data of the total $\bar np$ cross sections~\cite{Iazzi2000} and the ten level shifts and widths of the antiprotonic hydrogen of Refs.~\cite{Augsburger99} and \cite{Gotta99}.
The best fit to this data set yields a $\chi^2/\mbox{data}=4.52 $ for the present updated $N\bar N$ Paris potential.
Here, the $\chi^2/\mbox{data}$ are calculated as in Ref.~\cite{Paris94}.
The Paris 99 version~\cite{Paris99} has a corresponding $\chi^2/\mbox{data}=4.59$.
The values of the 15 parameters obtained in the present fit are compared to those of Paris 99 in Table~\ref{potparam}.

\begin{table*}
\caption{Results of the fit to the level shifts $\Delta E_L$ and width $\Gamma_L$ of antiprotonic hydrogen for the present work compared to the experimental data.
Units of $\Delta E_L$ and  $\Gamma_L$ are keV for $S$ waves and meV for $P$ waves.
Results for the Paris 99 potential~\cite{Paris99} are predictions.
The corresponding Coulomb corrected scattering lengths $a_c^L$ are obtained from Eq.~(\ref{DeltaES}) for $S$-waves with a principal quantum number $n=1$ and from Eq.~(\ref{DeltaEP}) for $P$-waves $(n=2)$.
One has $a(p\bar p)=[a(\mbox{T=0})+a(\mbox{T=1})]/2$, $a(S\mbox{-world})=[a(\mbox{singlet})+3a(\mbox{triplet})]/4$ and $a(\mbox{Sum-}P)=[3a(^1P_1)+3a(^3P_1)+5a(^3P_2)]/11$.
We use here the standard spectroscopic notation $^{2S+1}L_J$ for a given partial wave of spin $S$, of angular momentum $L$ and total angular momentum $J$. \label{levelshifts}}
\begin{ruledtabular}
\begin{tabular}{c|lll|lll}
            & \multicolumn{3}{c|}{$\Delta E_L-i\Gamma_L/2$}  & \multicolumn{3}{c}{$a_c^L\ [\mbox{fm}^{2L+1}]$}\\
 \hline
State   & Experimental & Present work & Paris 99 & Experimental & Present work & Paris 99 \\
$^1S_0$   & 0.440(75)-i0.60(12)~\cite{Augsburger99}  & 0.778-i0.519 & 0.755-i0.243  & 0.492(92)-i0.732(146) & 0.920-i0.666 & 0.911-i0.312 \\
$^3S_1$   & 0.785(35)-i0.47(4)~\cite{Augsburger99}  & 0.693-i0.393 & 0.654-i0.323  & 0.933(45)-i0.604(51) & 0.823-i0.498 & 0.778-i407 \\
$S$-world &  0.712(20)-i0.527(33)~\cite{Augsburger99} & 0.714-i0.425 & 0.680-i0.303  & 0.835(25)-i0.669(42) &0.847-i0.540 & 0.812-i0.384 \\
$^3P_0$ & -139(30)-i60(12)~\cite{Gotta99} & -67.0-i60 & -68.0-i66.8  & -5.68(1.23)-i2.45(49) & -2.74-i2.460 & -2.78-i2.730 \\
Sum-$P$ & 15(25)-i15.2(1.5)~\cite{Gotta99} & 6.10-i21.7 & 4.40-i10.9  & 0.613(1.02)-i0.621(60) & 0.250-i0.886 & 0.180-i0.445 \\
$^1P_1$ &        & -29.4-i13.2 & -29.6-i13.7 & &  -1.20-i0.539 & -1.21-i0.561 \\
$^3P_1$ &        & 63.8-i44.8 & 59.7-i12.6 & & 2.61-i1.83 & 2.44-i0.516 \\
$^3P_2$ &        & 7.22-i12.9 & -8.44-i8.12 & & -0.295-i0.528 & -0.345-i0.332
\end{tabular}
\end{ruledtabular}
\end{table*}

\begin{figure*}[ht]
\includegraphics[scale=0.5]{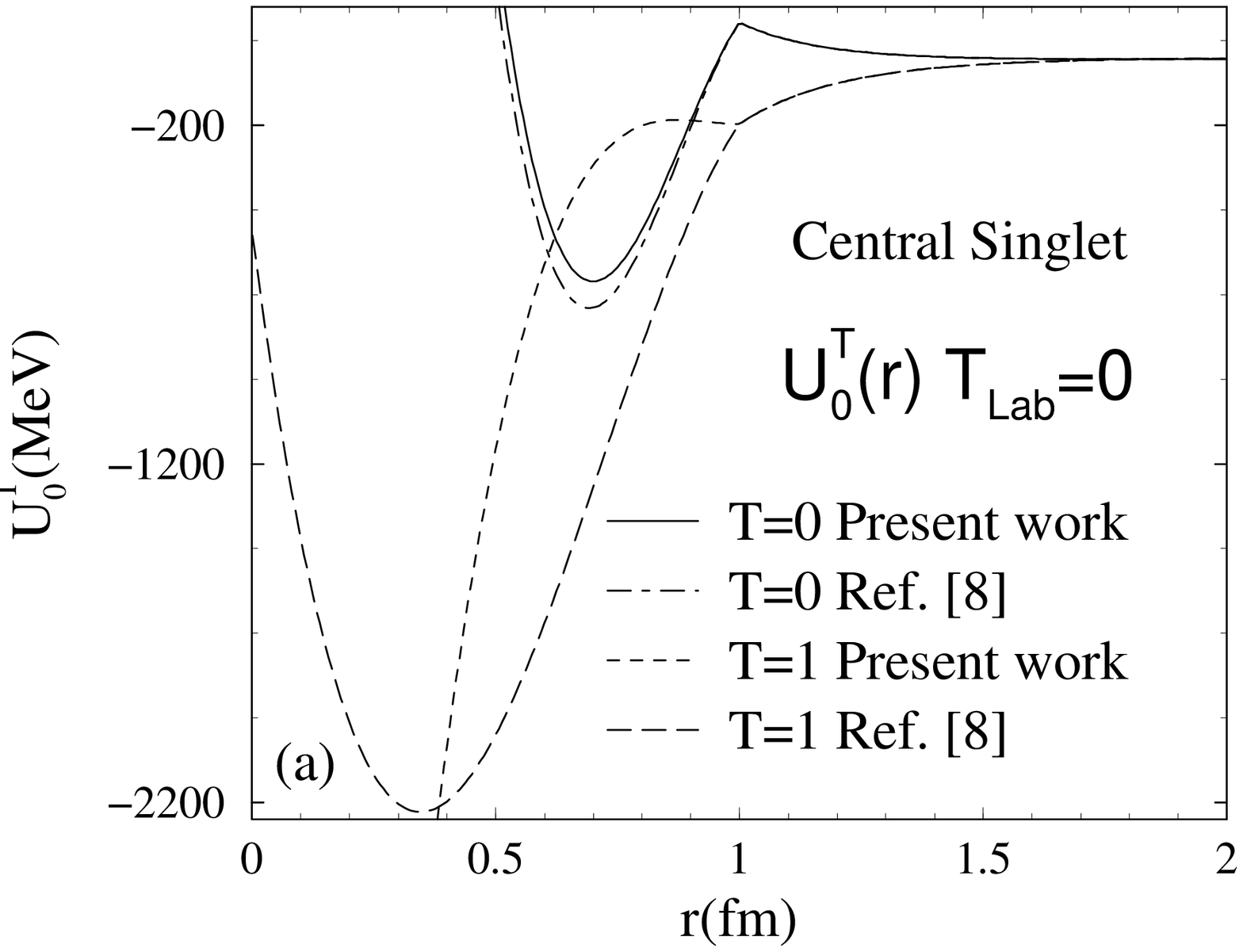}
\includegraphics[scale=0.5]{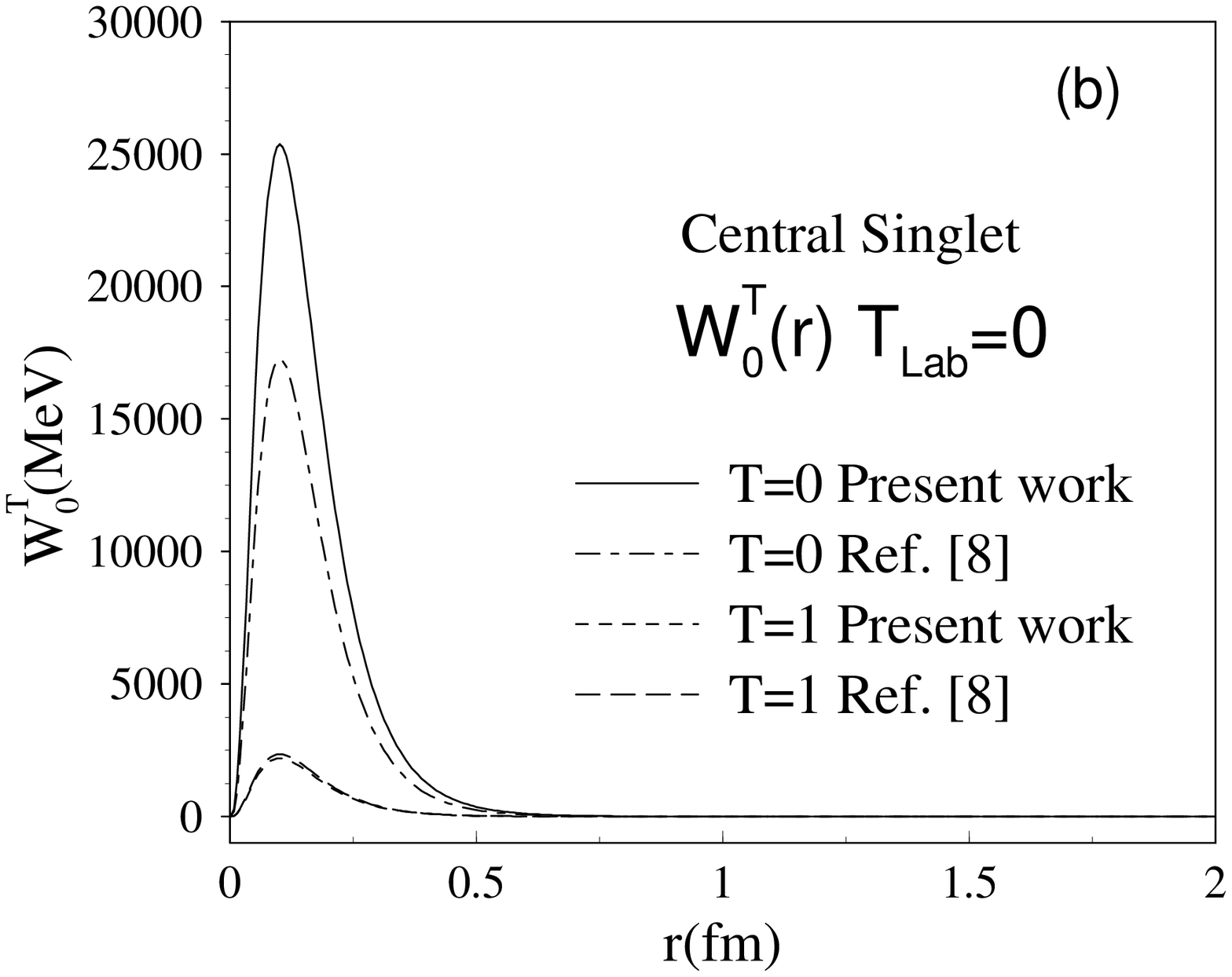}
\includegraphics[scale=0.5]{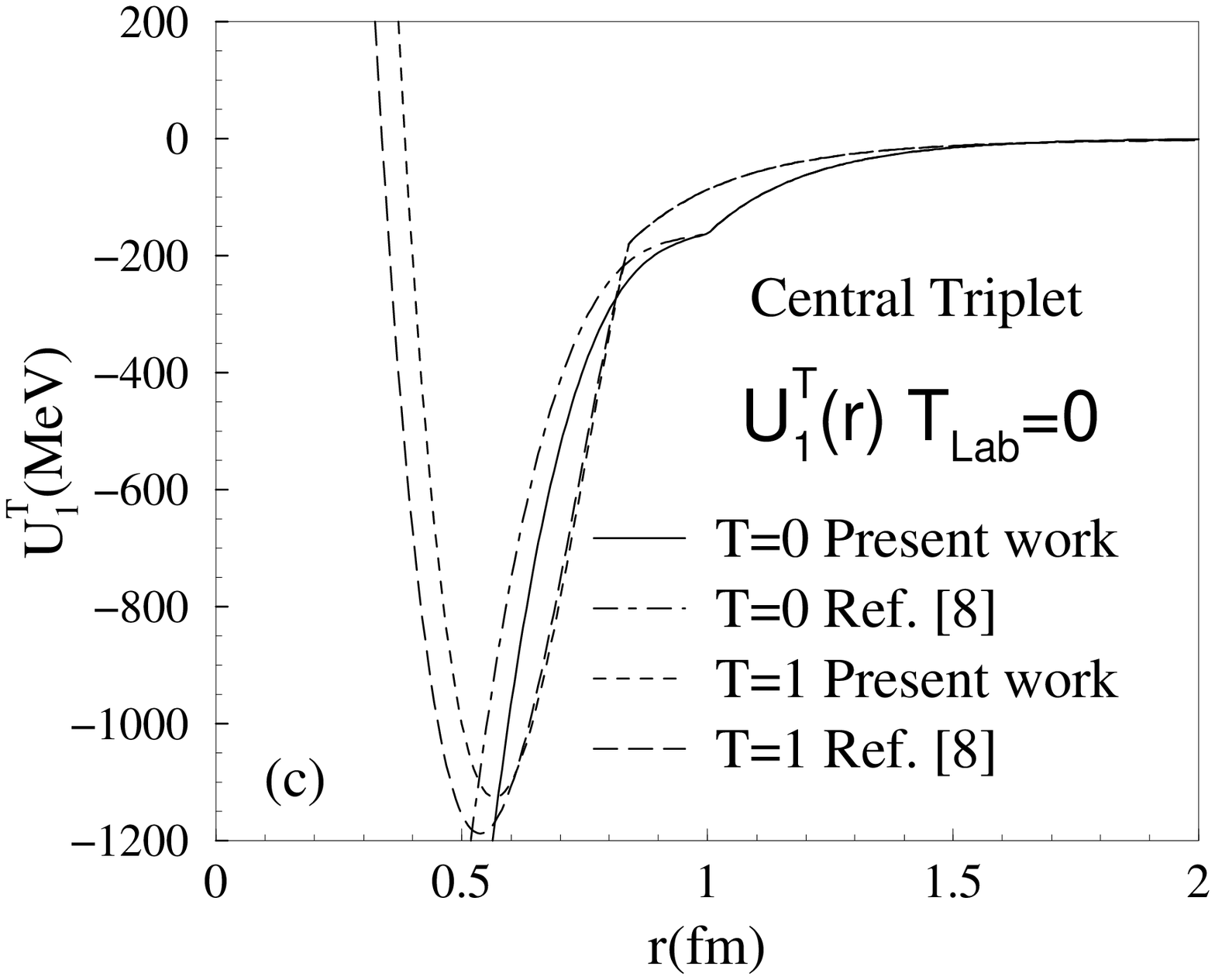}
\includegraphics[scale=0.5]{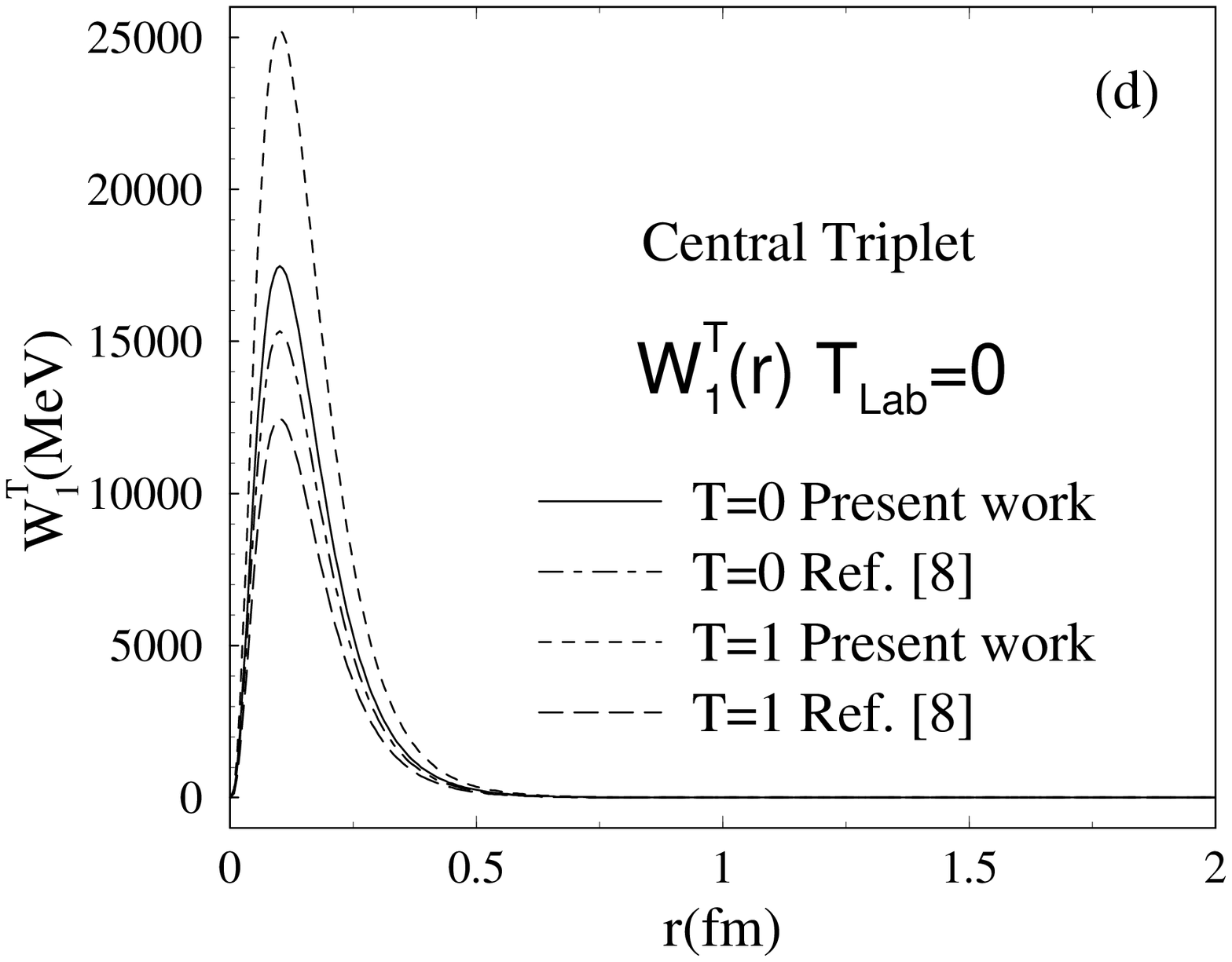}
\caption{\label{usinglet}Resulting  central real $U_{N\bar N}(r,0)$ and imaginary $W_{N\bar N}(r,0)$ potentials [see Eq.~(\ref{opticalpotential})] compared with those of Paris 99~\cite{Paris99}. 
The detailed definition of these potential can be found in the Appendix.}
\end{figure*}

\subsection{Scattering observables \label{scattering}}

In more detail, the value of $\chi^2$/data are 6.66(9.76) for the 106 $\bar pp$ total cross section data (see Fig.~\ref{Fig1}(a)), 2.66(4.47) for the 48 $\bar pp$ annihilation cross section data (see Fig.~\ref{Fig1}(b)), 2.03(4.29) for the 64 recent $\bar np$ total cross section data of Ref.~\cite{Iazzi2000} (see Fig.~\ref{Fig1}(c)),  0.87(1.56) for the 46 $\bar np$ annihilation cross section data (see Fig.~\ref{Fig1}(d)),  2.29(3.08) for the 28 $\bar pp$ backward elastic differential cross section data of Alston Garnjost \textit{et al.} (see Ref.~[6] in Ref.~\cite{Paris94} and Fig.~\ref{alston}(a)), 4.17(4.05) for the 3392 $\bar pp$ elastic data (see  more elastic data in Figs.~\ref{alston}(b-d) and 6.74(6.84) for the 639 $\bar pp\to\bar nn$ charge exchange data (see some examples in Fig.~\ref{cex}).
The $\chi^2/\mbox{data}$ are 6.71(11.75) for the ten level shifts and widths for the antiprotonic-hydrogen data~\cite{Augsburger99,Gotta99} (see Table~\ref{levelshifts}).
All the above $\chi^2/\mbox{data}$ quoted in parenthesis correspond to  calculations with the Paris 99 potential~\cite{Paris99}.
Significant improvements were obtained with the Paris 99 potential compared to the earlier Paris 82~\cite{Paris82} and 94~\cite{Paris94} versions.
As seen from the above detailed $\chi^2$ and in comparison to the Paris 99 version~\cite{Paris99}, the present potential yields an improved description of all experimental observables  but  the  $\bar pp$ elastic data.

The results of the fit to the total and annihilation $\bar pp$ and $\bar np$ cross sections are shown in Figs.~1(a-d).
For $T_{Lab}\lesssim$100~MeV the present model is closer to the data than the Paris 99 potential.
The recent $\bar np$ total cross section (Fig.~\ref{Fig1}(c)) is well reproduced.
One has (see Fig.~\ref{alston}(a))  a better description of the backward elastic differential cross sections for $T_{Lab}\gtrsim$130~MeV.
A sample of $\bar p p$ elastic differential cross section at 288.3 MeV (Fig.~\ref{alston}(b))  and polarization at 146.3  (Fig.~\ref{alston}(c)) and 219.9 MeV  (Fig.~\ref{alston}(d)) are displayed in  Fig.~\ref{alston}.
As mentioned above, the present model is not as good as the Paris 99 version on $\bar p p$ elastic data.
Despite a slight improvement on the total $\chi^2$/data for the $\bar p p\to \bar n n$ charge-exchange, CEX, data (see above), it can be seen in Fig.~\ref{cex}  that both potentials fail to give a good fit.
Had we allowed a 0.85 normalization factor the reproduction of the integrated CEX cross section (see Fig.~\ref{cex}(a)) would have been better.
Note that at 147.3 MeV (Fig.~\ref{cex}(b)) a normalization factor of 0.74 is needed to reproduce the data.
None of the two versions  fit well the CEX polarization data as seen for instance in Fig.~\ref{cex}(c) and~(d).

\subsection{Antiprotonic hydrogen level shifts \label{shifts}}

\begin{table}
\caption{Binding energy in MeV of the close to threshold quasibound states of the present model and of the Paris 99 potential~\cite{Paris99}.\label{bindings}}
\begin{ruledtabular}
\begin{tabular}{ccc}
$^{2T+1\ 2S+1}L_J$ & Present work & Paris 99 \\
\hline
$^{11}S_0$ & -4.8-i26 & \\
$^{33}P_1$ & -4.5-i9.0 & -17-i6.5
\end{tabular}
\end{ruledtabular}
\end{table}

\begin{table}
\caption{Close to threshold resonances of the present model.
The numbers in parenthesis correspond to the $^{11}P_1$ resonance of Paris 99.
The $^{13}P_0$ and $^{13}P_1$ resonances have identical positions in the Paris 99 model.
There is no $^{33}P_0$ resonance in the Paris 99 potential. \label{resonances}
}
\begin{ruledtabular}
\begin{tabular}{ccccc}
$^{2T+1\ 2S+1}L_J$ & $^{11}P_1$ & $^{13}P_0$ & $^{13}P_1$ & $^{33}P_0$ \\
\hline
Mass (MeV) & 1877 (1872) & 1876 & 1872 & 1871 \\
Width (MeV) & 26 (12) & 10 & 20 & 21 \\
\end{tabular}
\end{ruledtabular}
\end{table}

\begin{figure*}[ht]
\includegraphics[scale=0.5]{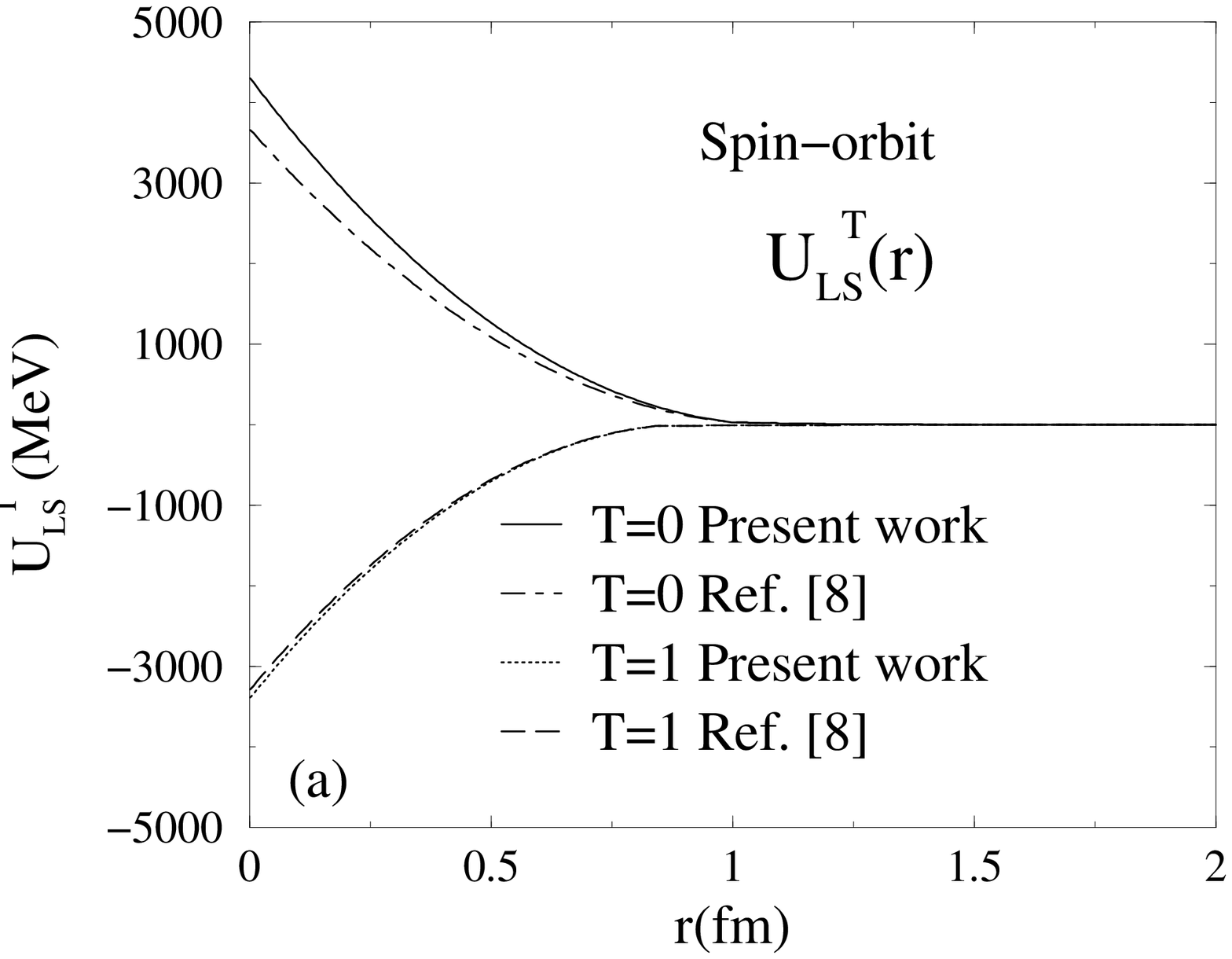}
\includegraphics[scale=0.5]{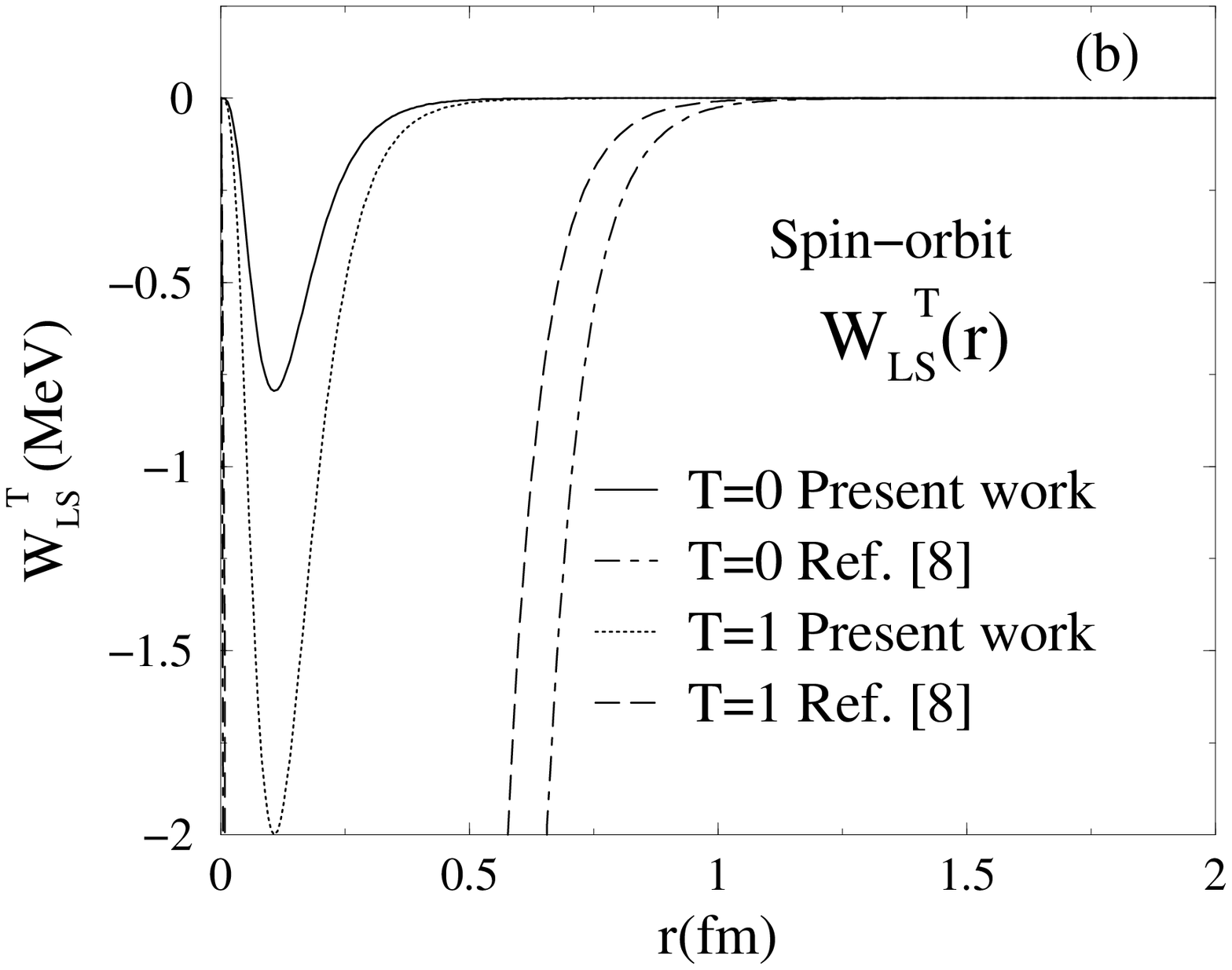}
\includegraphics[scale=0.5]{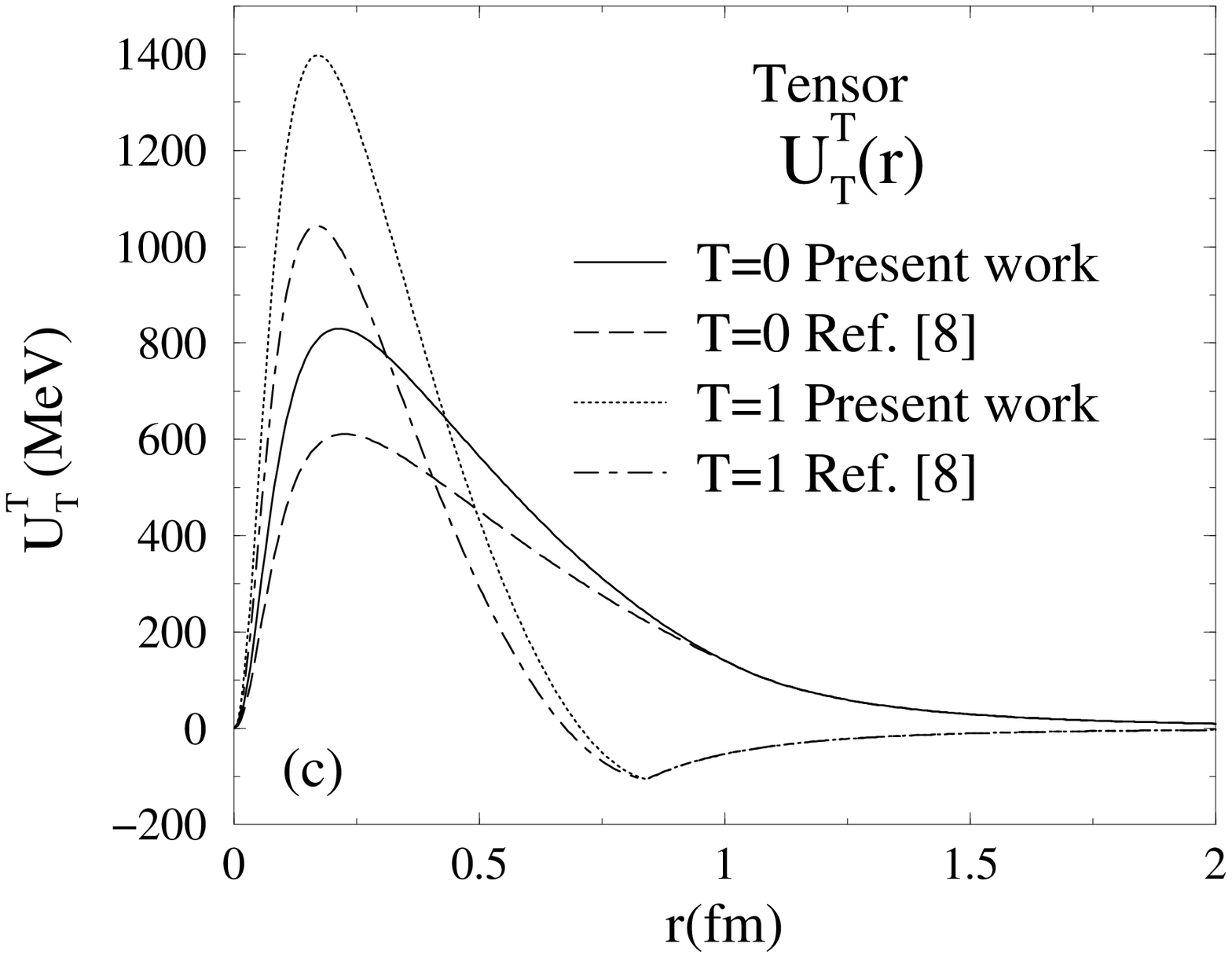}
\includegraphics[scale=0.5]{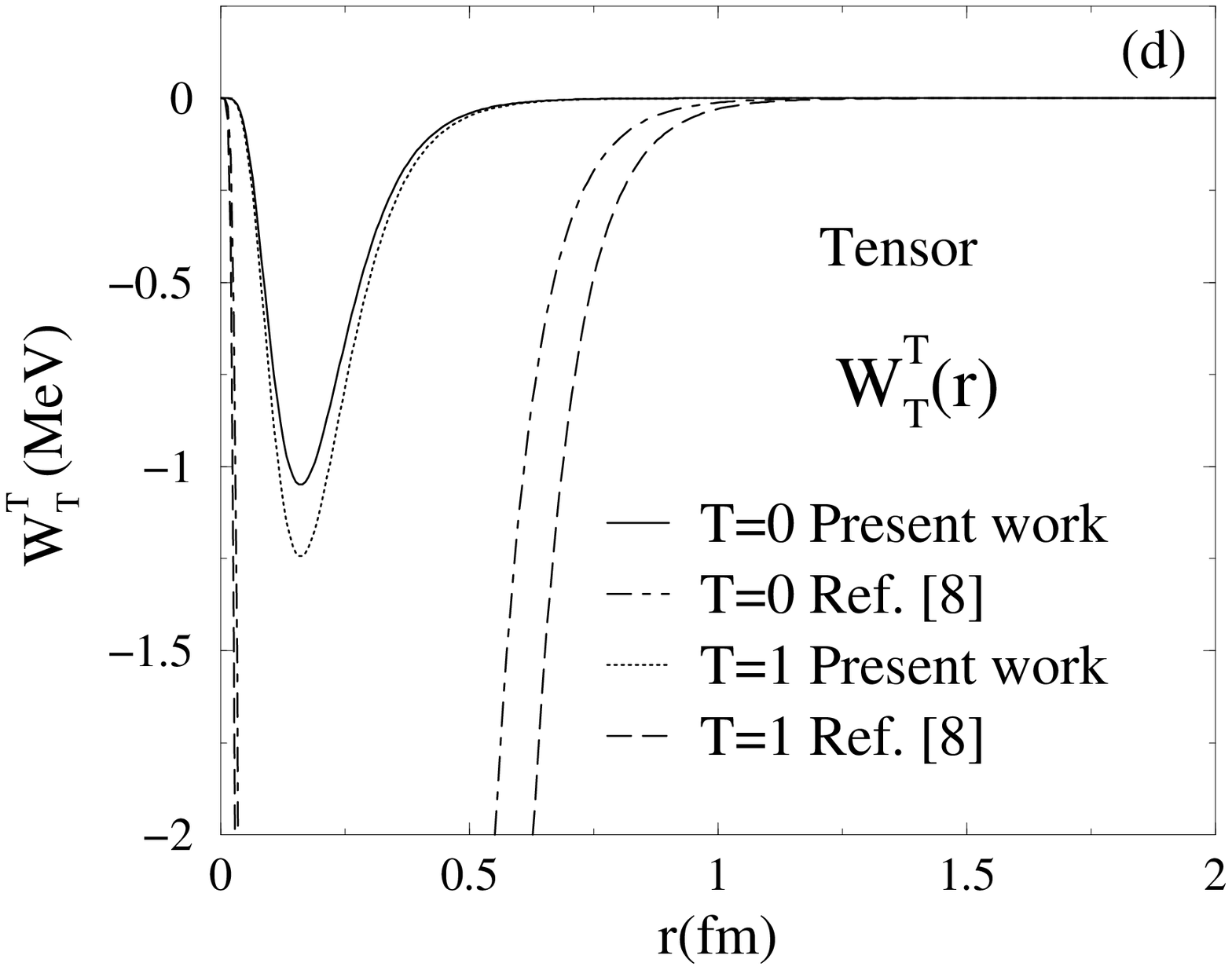}
\caption{\label{Figlst} As in Fig.~\ref{usinglet} but for the spin-orbit and tensor optical potentials.}
\end{figure*}

A comprehensive description on the protonium is displayed in Ref.~\cite{Klempt2002}.
Here, the $S$-wave atomic level shifts $\Delta E_S$ and widths $\Gamma_S$ are related to the Coulomb corrected $S$-wave $p\bar p$ complex scattering length $a_c^S$ by
(see Eq.~(3.7) of Ref.~\cite{Trueman61}),
\begin{equation}
\label{DeltaES}
\Delta E_S-i\frac{\Gamma_S}{2}=\frac{2\pi}{\mu_{p\bar p}}
\left\vert
\psi^{\mbox{coul}}(0)
\right\vert^2
\ a_c^S
\left(
1-3.154\ \frac{a_c^S}{B}
\right).
\end{equation}
This relation is accurate to second order in $a_c^S/B$, good enough for the $p\bar p$ system.
In Eq.~(\ref{DeltaES}) $\mu_{p\bar p}=M_P/2$ is the $p\bar p$ reduced mass, $M_P=938.27$~MeV being the proton mass.
The Bohr radius is $B=1/(\alpha\mu_{p\bar p})=57.6399$~fm,  where $\alpha=1/137.036$ is the fine structure constant.
The modulus of the zero range Coulomb wave function $\psi^{\mbox{coul}}(0)$ is given by
$\left\vert
\psi^{\mbox{coul}}(0)
\right\vert^2=1/(\pi B^3)$.
It can be seen that for a scattering length of the order of 1 fm, the second term of Eq.~(\ref{DeltaES}) is a correction of a few per cent.
Such a correction is negligible in higher angular momentum states and for $P$-wave ($n=2$) we use~\cite{Lambert70}
\begin{equation}
\label{DeltaEP}
\Delta E_P-i\frac{\Gamma_P}{2}=\frac{3}{16\mu_{p\bar p}B^5}\ a_c^P.
\end{equation}
We follow Eqs.~(4) to (6) of Ref.~\cite{Carbonell92} to
calculate the Coulomb corrected scattering lengths. The phase
shifts are obtained by solving the $p\bar p$ Schr\"odinger
equation in configuration space with the Paris $N\bar N$ optical
potential plus the Coulomb potential.

The resulting $S$- and $P$-wave antiprotonic-hydrogen level shifts and widths are compared to the experimental ones~\cite{Augsburger99,Gotta99} in Table~\ref{levelshifts}.
There is an overall improvement in comparison with the predictions of the Paris 99 potential.
Furthermore, our predictions of $^1P_1,\ ^3P_1$ and $^3P_2$ waves are given in this table.
We also list all the corresponding Coulomb corrected scattering lengths related here to the level shifts and widths through Eqs.~(\ref{DeltaES}) and (\ref{DeltaEP}) for $S$ and $P$ waves, respectively.
Some of these numbers can be compared to those given in Table III of Ref.~\cite{Entem2006}, where
a $N\bar N$ potential is derived from a quark-based $NN$ model~\cite{Entem2000} supplemented by a state independent, phenomenological imaginary potential of Gaussian type.
Energy shifts and widths show differences.
All models give a value of $\Delta E(^3P_0)$ about half of the experimental value.
Due to the dominance of the one pion exchange for the $^1P_1$ state,
these models have similar results for
$\Delta E(^1P_1)$ and $\Gamma(^1P_1)$.

\subsection{Bound states and resonances \label{boundstates}}

We search for
the close to threshold bound states 
or resonances
present in our model and in the Paris~99 potential~\cite{Paris99}.
As seen in Figs.~\ref{usinglet}a and~\ref{usinglet}c, the real central singlet and triplet potentials have relatively strong medium and short range attractive parts.
If the imaginary potentials are set to zero, several bound or resonant states exist, however many of them disappear when the necessary annihilation is introduced~\cite{Lacombe84}. 
Following the method of this last reference, 
we find an isospin $T$=0 $S$-wave quasibound state $^{11}S_0$ (using the notation $^{2T+1\  2S+1}L_J$) of 4.8 MeV binding and of 52 MeV total
width  (see Table~\ref{bindings}). Such a state is absent in the Paris 99~\cite{Paris99} , Paris
94~\cite{Paris94} and Paris 82~\cite{Lacombe84} potentials.
A 9 MeV bound  $^{33}P_1$ state of 18 MeV total width is found.
This triplet $P$-wave state is also found in the Paris 
99 (see Table~\ref{bindings}) and Paris 94 (see Table VI of Ref.~\cite{Paris94})
models. The well known $^{13}P_0$ resonance, resulting from an
attractive one-pion exchange force, is present in all potentials (see Table~\ref{resonances}).
This state has also been found in the recent $N \bar N$ constituent
quark model of Ref.~\cite{Entem2006}. These close to threshold
quasibound states and resonances are very difficult to detect in
$N \bar N$ scattering data as, right from threshold, many partial
waves contribute (see for instance the Paris 94 phases in
Ref.~\cite{Paris94}). Chances to observe these states are larger
in reaction processes which select the partial wave contributions
such as in $J/\Psi$ decays~\cite{Loiseau2005,Dedonder2008}.

\subsection{The  $N\bar N$ optical potential \label{potential}}

The values of the short-range parameters of the potential listed in Table~\ref{potparam} together with the plots in Figs.~\ref{usinglet}(a-d) of the central optical potentials at threshold ($T_{Lab}=0$) show the differences between the present potentials and those of 1999~\cite{Paris99}.
The isospin 0 real central singlet potentials are very similar (see Fig.~\ref{usinglet}(a)).
The isospin 1 real central singlet potential, less attractive for $r\gtrsim 0.5$ fm than that of Paris 99, has a much deeper short range part (see Fig.~\ref{usinglet}(a)) and compare the corresponding $U_0^a(r_3)$ and $U_0^a(r_2)$ in Table~\ref{potparam}.
The imaginary $T=0$ central singlet potential is more repulsive than
that of Paris 99 (see Fig.~\ref{usinglet}(b)) and its energy dependence is different as can be seen from the corresponding $f$ values in Table~\ref{potparam}.   
The present imaginary $T=1$ singlet potential, close to that of Paris 99 (see Fig.~\ref{usinglet}(b)) is weak, however both potentials differ in  their energy dependence (compare their $f$ values in Table~\ref{potparam}).
The real central triplet potentials are quite attractive (see Fig.~\ref{usinglet}(c)) which supports the $^3P$-wave quasibound states and resonances listed in Tables~\ref{levelshifts} and III.
The imaginary $T=0$ and $T=1$ triplet potentials are more repulsive than
those of Paris 99 (see Fig.~\ref{usinglet}(d)).
The short range real spin-orbit potentials (Fig.~\ref{Figlst}(a)) are very similar in both models while those of the tensor are stronger in the present model  (Fig.~\ref{Figlst}(c)).
The imaginary spin-orbit  (Fig.~\ref{Figlst}(b) and tensor  (Fig.~\ref{Figlst}(d) are much smaller (compare also
the $g_{LS}$ and $g_T$ given in Table~\ref{potparam}).

\section{Discussion \label{discussion}}

As can be seen in Figs.~\ref{usinglet}a, \ref{usinglet}c, and \ref{Figlst}c  there is a more or less sharp  change  at the matching radius $r=r_c$  with $r_c=0.84$ or 1~fm (see the Appendix) between the phenomenological part  of the potential and the theoretical part  deduced from the Paris $NN$ potential.
This variation is not  due to the real part of the annihilation potential which in our model is very short 
ranged.
As reminded in the Appendix, the $N \bar N$ annihilation into mesons, with intermediate nucleon-antinucleon state in the crossed channel, has been  shown by dispersion techniques in Ref.~\cite{Paris94} to lead to a very short ranged absorptive potential. 
The real part due to annihilation is expected to be also very short ranged. 
The high-partial wave analysis of the 
$NN$ interactions performed  in Ref.~\cite{RVM73} with the Paris model
shows indeed that there is some limit of applicability for the $G$ transformed
potential.
The uncertainty of the meson exchange interaction increases while its
range becomes smaller.
Then, in the 1 fm range and below,
the real phenomenological short ranged part is attributed mainly 
to two sources: (i) to exchange of heavier mesons other than those 
explicitly included in the  model, viz. $\pi$, 2$\pi$, $\omega$ and  $A_1$,
(ii)  to involved quark-antiquark (and gluon) forces based on QCD.
 The relatively sharp junctions between the real theoretical medium ranged potentials and the real phenomenological short ranged ones might then indicate some of these uncertainties in  the $G$-parity transformed meson-exchange forces. 

The present updated Paris potential gives a better description 
of all observables but the $\bar p p$ elastic data.  
As noted above in Sec.~\ref{scattering} it has only a slightly better  overall $\chi^2$/data of 4.52 versus  4.59 (Paris 99) for the 4333 data considered. 
However the low energy data, $T_{Lab} \lesssim$~50~MeV, including 
the atomic data are better described (see below).
We do not pretend that the present model is a better one but it is a different and an interesting  one compared to the 1999 version in the sense that it generates  a close to threshold quasibound  
$^{11}S_0$ state.

It has to be stressed that the actual contribution of the $^{11}S_0$ partial wave to the cross sections is small by the statistical reasons. 
In this sense the parameters of this wave cannot be determined  precisely.
This statement is mainly valid for $T_{Lab} \gtrsim$~50~MeV when the $P$-wave contributions start to be very large.   
The atomic level shift and width in the spin singlet state are more constraining.
However, even there,  the contribution of the spin-0 state is only 1/3 of the contribution from the spin-1 state and the total contribution of the $^{11}S_0$  wave to the spin singlet atomic shift is about 25\%. 
Fortunately,  the new $\bar n p$ scattering data  (which involves $T=1$ amplitudes) allow one  to fix the  isospin 1 contribution in a better way. 
In particular the absorptive parts of scattering amplitudes are larger.
It is the consistency of  scattering data and of the atomic widths that induce the energy dependence of absorptive potential  $W_{N \bar N} (\mathbf{r},T_{Lab})$ to differ from the previous models.  
This reflects upon the position  of the  $^{11}S_0$ state. 
In all Paris models,  the potential in  $^{11}S_0$ state has been strongly attractive.  
Deeply bound  states have been formed with binding energies in the few hundred MeV range. 
These states are not realistic as the extrapolation of the $W_{N \bar N} (\mathbf{r},T_{Lab})$ so far down below the threshold is very uncertain.  
It is the next state which may or may not exists close to the threshold that is of our interest. 
It does not exist in Paris 99 version and arises in the present Paris one.  
The main difference that creates it,  is  the energy dependence of 
the $W_{N \bar N} (\mathbf{r},T_{Lab})$  generated by the $\bar n p$ data and the atomic level 
widths.

For completeness, we show in Table~\ref {a1S0nc} the spin-singlet (not corrected for Coulomb) scattering lengths in both models. 
The increase of absorption is visible,  Im~$a(^{11}S_0)$ is determined 
fairly  precisely  as both   Im~$a(^{31}S_0)$ and   Im~$a(^{1}S_0)$ are now 
better known from the~$\bar n p$ data and the antiprotonic hydrogen atom widths, respectively. 
It is not the case with the real parts  which are larger than  
those obtained from atoms as seen in Table~\ref{levelshifts}. 
On the other hand 
Figs.~\ref{Fig1}a and \ref{Fig1}b  indicate that our total and annihilation $\bar p p$ cross sections are too small at $T_{lab} \lesssim 50$ MeV. 
It shows some inconsistency, but it is also clear that the present potential is doing much better than Paris 99 in this low energy and atomic region. 
This  inconsistency does not necessarily reflect on 
the existence of the quasibound state.
There are two reasons for that, one being, as mentioned above, the low statistical weight of the $^{11}S_0$ state. 
The other is the fact that Re~$a(^{11}S_0)$ is not 
easily related to the bound state energy. 
Due to the peculiar shape of  central singlet potential (Fig.~\ref{usinglet}(a), the effective range expansion has a very short convergence radius and one 
cannot infer the binding energy from the scattering length. 

Let us also add that we have not been motivated to obtain 
the quasibound state. 
We do not sacrifice the low energy data at the expense of atomic data. 
The best fit gives an  ``in between" solution. 
Note that the partial values of $\chi^2$ for data limited to $T_{lab} \lesssim 50$ MeV  with the atomic data  (altogether 413 data) are  3544 for the present 
model and 5218 for the Paris 99 solution. 
A sizable fraction of these  $\chi^2$ comes from charge exchange reactions.  
In fitting, we  obtain several almost 
equivalent solutions as far as the minimum $\chi^2$ is concerned, although 
the present model has the best overall $\chi^2$. 
These alternative solutions did not always produce  the $^{11}S_0$ quasibound state. 

\begin{table}
\caption{The spin-singlet (not corrected for Coulomb)
scattering lengths of the present model and of the Paris 99 potential~\cite{Paris99}.\label{a1S0nc}}
\begin{ruledtabular}
\begin{tabular}{ccc}
$a(^{2T+1\ 2S+1}S_J)$ & Present work & Paris 99 \\
\hline
$a(^{31}S_0)$ & 0.684-i 0.473 & 0.979 -i0.294\\
$a(^{11}S_0)$ & 1.115-i 0.856 & 0.844-i0.329
\end{tabular}
\end{ruledtabular}
\end{table}

\section{Summary and conclusions \label{summary}}

We have redetermined the short-range parameters of the Paris $N\bar N$ optical potential by fitting, besides the set of data used in the previous 1999 version, recent antiprotonic-hydrogen level shifts and widths and total $\bar np$ cross sections.
Improvements in the fit are obtained.
This model predicts quasibound states close to the threshold in the $p\bar p(^{11}S_0)$ and $p\bar p(^{33}P_1)$ waves.
Existence of these states indicates a strong dependence on the parameters of the model.
There is also a well established resonance in the $p\bar p(^{13}P_0)$ wave.
The $^1S_0$ state reproduces well~\cite{Loiseau2005} the recent $\gamma p\bar p$ spectrum measured by the BES Collaboration~\cite{Bai2003}.
The observed peak in the invariant mass of the produced mesons in the $J/\psi \to \gamma \pi^{+}\pi^{-}\eta'$ decay~\cite{Ablikim2005} can be explained by an interference of the quasibound state $^{11}S_0$ with a background amplitude~\cite{Dedonder2008} .
Let us mention that for uses of the Paris $N \bar N$ potential, like for the initial or final state interactions in various processes, the present potential can be provided upon request.

\begin{acknowledgments}
We acknowledge useful discussions on quasibound states and resonances with B.~Moussallam. 
We also thank J.-P.~Dedonder and O. Leitner for helpful comments. 
M.L. and B.L. are grateful for valuable exchanges with Yupeng Yan. 
This work was supported in part  by the Department of Energy, Office of Nuclear Physics, Contract  No. DE-AC02-06CH11357. 
This research was also performed in the framework of the IN2P3-Polish Laboratory Convention (collaboration No. 05-115).
\end{acknowledgments}

\appendix
\section{Full expression of the optical potential\label{appendix}
 }

For completeness, the full expression of the optical potential is revisited below.
For each isospin value $T=0$ or $T=1$, the real potential can be expressed in terms of the five usual nonrelativistic invariants,
 $\Omega_0=\left(1-\mbox{\boldmath $\sigma$}_1\cdot \mbox{\boldmath $\sigma$}_2\right)/4$, 
 $\Omega_1=\left(3+\mbox{\boldmath $\sigma$}_1\cdot \mbox{\boldmath $\sigma$}_2\right)/4$, 
 $\Omega_{LS}=\mathbf{L}\cdot\mathbf{S}$, 
  $\Omega_T=
  3(\mbox{\boldmath $\sigma$}_1\cdot \mathbf{r}\ \mbox{\boldmath $\sigma$}_2\cdot \mathbf{r})/{r^2}
  -\mbox{\boldmath $\sigma$}_1\cdot \mbox{\boldmath $\sigma$}_2$ 
  and 
  $\Omega_{SO2}=(\mbox{\boldmath $\sigma$}_1\cdot\mathbf{L}\ \mbox{\boldmath $\sigma$}_2\cdot \mathbf{L}
  +\mbox{\boldmath $\sigma$}_2\cdot \mathbf{L}\ \mbox{\boldmath $\sigma$}_1\cdot \mathbf{L})/2$. 
  It reads

\begin{eqnarray}
\label{realpotential}
U_{N \bar N}\left({\bf r}, T_{Lab})\right)  =&&U_0\left(r,T_{Lab}\right) \Omega_0 + 
U_1\left(r,T_{Lab}\right) \Omega_1 \nonumber \\
&+& U_{LS}(r) \Omega_{LS} + U_{T}(r) \Omega_{T} \nonumber \\
&+& U_{SO2}(r) \Omega_{SO2}.
\end{eqnarray}
The linear nonlocality in the central singlet, $U_0$ and central triplet, $U_1$ potentials is expressed as
\begin{equation}
 \label{centralpotential}
 U\left(r,T_{Lab}\right)= U^a(r)+T_{Lab}\ U^b(r).
\end{equation}
The potential, for $r \ge r_c\ (r_c \le 1$ fm), is the $G$-parity transform of the theoretical Paris $NN$ potential~\cite{Paris80}. 
We use, for the $\omega$ exchange ($m_\omega=782.7$ MeV), $g_\omega^2/4\pi=11.75$ as in Ref.~\cite{Paris80}.
However, in order to have a more attractive isospin 0 central singlet potential in the vicinity of 1 fm, we modify the coupling of the shorter range $A_1$ exchange ($m_{A_1}= 1100$ MeV) from $g^2_{A_1}=14.$ to 10.4.

For $r \le r_c$, the empirical potentials are given by a cubic $r$ expression for the central $U^a_0(r)$ and  $U^a_1(r)$ terms:
\begin{equation}
U(r) =a_3r^3+a_2r^2+a_1r+a_0
\label{cubic}
\end{equation}
and by a quadratic one for, $U_{0,1}^b(r)$, $U_{LS}(r)$, $U_T(r)$ and $U_{S02}(r)$:
\begin{equation}
U(r) =b_2r^2+b_1r+b_0.
\label{quadratic}
\end{equation}
The parameters $a_i$ ($i= 0$ to 3) and $b_i$ ($i=0$ to 2) are determined (i) by matching to the theoretical potential at $r=r_0= r_c$ and $r=r_1=r_0 +\Delta r$ with $\Delta r=0.15$~fm, (ii) choosing a phenomenological height at $r_2=0.587$~fm and at $r=0.188$~fm  (here only for $U_{0,1}^a(r)$).
For all isospin-0 potentials $r_c=1$~fm and for all isospin-1 terms $r_c=0.84$~fm except for $U_0^a(r)$ where $r_c=1$~fm.
While solving the Sch\"odinger equation we have regularized the tensor potential $U_T(r)$ at small $r$ by multiplying it by
\begin{equation}
F(r) =\dfrac{(pr)^2}{1+(pr)^2},
\label{regUT}
\end{equation}
with $p=10$ fm$^{-1}$.

As shown in the Appendix of Ref.~\cite{Paris94}, the imaginary potential $W_{N\bar N} (\mathbf {r}, T_{Lab})$, arising from nucleon-antinucleon annihilation into mesons, can be approximated by a short range radial function proportional to effective phenomenological couplings with a linear energy dependence for the central and spin-spin components. One writes
\begin{eqnarray}
W_{N\bar N} \left( \mathbf {r}, T_{Lab} \right)=  &\Bigg[&  g_C \left(1+f_CT_{Lab} \right)  \nonumber   \\  
& + & g_{SS}\left(1+f_{SS}T_{Lab}\right) \mbox{\boldmath $\sigma$}_1 \cdot \mbox{\boldmath $\sigma$}_2 + g_T \Omega_T
 \nonumber   \\  
  &+&\dfrac {f_{LS}} {4m^2} \Omega_{LS} \dfrac {1} {r} \dfrac {d} {dr} \Bigg{ ] }
  \dfrac {K_0(2mr)} {r},
 \label{WNNBrT}
\end{eqnarray}
where the modified Bessel function $K_0(2mr)$ is the Fourier transform of a dispersion type integral resulting from the calculation of the $N\bar N $ annihilation box diagram into two mesons with a nucleon-antinucleon intermediate state in the crossed $t$-channel~\cite{Paris94}. 
One has

\begin{equation}
K_0(2mr) =\dfrac{1}{2} \int_{4m^2}^\infty dt' \dfrac{e^{-\sqrt{t'}r}}{\sqrt{t'(t'-4m^2)}}.
\label{K0}
\end{equation}
In Eqs.~(\ref{WNNBrT}) and (\ref{K0}), $m$ is taken to be quite close to the nucleon mass, $m=940$~MeV.
To avoid the singular behavior at $r=0$, we regularize the central and spin-spin potential of Eq.~(\ref{WNNBrT}) by multiplying them with
\begin{equation}
G(r) = (1-e^{-2mr})^4.
\label{Gr}
\end{equation}
The imaginary spin-orbit and tensor potentials are  multiplied by
\begin{equation}
H(r) = (1-e^{-2mr})^7.
\label{Hr}
\end{equation}

The values of the empirical real potentials at $r=r_2$ and $r_3$ and of the parameters $g_i$ ($i=C$, $SS$, $T$ and $LS$) and  $f_i$ ($i=C$, $SS$), determined by the fit,  are displayed in Table~\ref{potparam}.
These 15 parameters play a different role in the fit.
The important parameters are \\(i)  the six values of the central singlet and triplet, of the tensor and of the $L\cdot S$ components at  $r=r_2=0.587$~fm for the real part,\\ (ii) the four couplings of the singlet and triplet central terms for the imaginary part.\\
A fine tuning of the fit is then obtained by adjusting the five remaining parameters.



\begin{thebibliography}{99}

\bibitem{Bai2003}
J. Z. Bai \textsl{et al.} (BES Collaboration),
Phys. Rev. Lett. \textbf{91}, 022001 (2003),
\textit{ Observation of a near threshold enhancement in the $p\bar p$ mass spectrum from radiative $J / \psi\to\gamma p\bar p$ decays}.

\bibitem{Wei2007}
J.-T. Wei \textsl{et al.}
(Belle Collaboration),
Phys. Lett. \textbf{B659}, 80 (2008),
\textit{Study of  the decay mechanism for $B^+ \to p \bar p K^+$ and $B^+ \to p \bar p \pi^+$}.

\bibitem{Chen2008}
J.-H. Chen \textsl{et al.}
(Belle Collaboration),
Phys. Rev. Lett. \textbf{100}, 251801 (2008),
\textit{Observation of $B^0 \to p \bar p K^{*0}$ with a large $K^{*0}$ polarization}.

\bibitem{Abe2002}
K. Abe \textsl{et al.}
(Belle Collaboration),
Phys. Rev. Lett. \textbf{89}, 151802 (2002),
\textit{Observation of $\bar B^0 \to D^{(*)0} p \bar p$}.

\bibitem{Loiseau2005}
B. Loiseau and S. Wycech,
Phys. Rev. C \textbf{72}, 011001 (2005),
\textit{Antiproton-proton channels in $J/\psi$ decays}.

\bibitem{Paris82}
J. C\^{o}t\'e, M. Lacombe, B. Loiseau, B.~Moussallam, and R.~Vinh Mau,
Phys. Rev. Lett. \textbf{48}, 1319 (1982),
\textit{Nucleon-Antinucleon Optical Potential}.

\bibitem{Paris94}
M. Pignone, M. Lacombe, B. Loiseau, and R.~Vinh Mau,
Phys. Rev. C \textbf{50}, 2710 (1994),
\textit{Paris $N\bar N$ potential and recent proton-antiproton low energy data}.

\bibitem{Paris99}
B. El-Bennich, M. Lacombe, B. Loiseau, and R.~Vinh Mau,
Phys. Rev. C \textbf{59}, 2313 (1999),
\textit{Refining the inner core of the Paris $N\bar N$ potential}.

\bibitem{Iazzi2000} 
F. Iazzi \textsl{et al.}, OBELIX Collaboration,
Phys. Lett. \textbf{B475}, 378 (2000),
\textit{Antineutron-proton total cross section from 50 to 400~MeV/c}.

\bibitem{Augsburger99}
M. Augsburger \textsl{et al.},
Nucl. Phys. \textbf{A658}, 149 (1999),
\textit{Measurement of the strong interaction parameters in antiprotonic hydrogen and probable evidence for an interference with inner bremsstrahlung}.

\bibitem{Gotta99}
D. Gotta \textsl{et al.},
Nucl. Phys. \textbf{A660}, 1283 (1999),
\textit{Balmer $\alpha$ transitions in antiprotonic hydrogen and deuterium}.

\bibitem{Sibirtsev2005} 
A. Sibirtsev, J. Haidenbauer, S. Krewald, Ulf-G.~Mei\ss ner, and A.~W.~Thomas,
Phys. Rev. D \textbf{71}, 054010 (2005),  
{\it{Near-Threshold Enhancement in the $ p\bar{p}$ Mass Spectrum in $ J/\psi $ Decays.}}

\bibitem{Paris80}
M. Lacombe, B. Loiseau, J-M. Richard, R.~Vinh Mau, J.~C\^{o}t\'e,  P. Pir\`es, and R. de Tourreil,
Phys. Rev. C\textbf{21}, 861 (1980),
\textit{Parametrization of the Paris $NN$ potential}.
 
\bibitem{Cottingham73}
W. N. Cottingham, M. Lacombe, B. Loiseau, J.-M.~Richard, and R.~Vinh Mau,
Phys. Rev. D\textbf{8}, 800 (1973),
\textit{Nucleon-Nucleon interaction from pion-nucleon phase shift analysis}.

\bibitem{Klempt2002}
 E. Klempt, F. Bradamante, A.~Martin and J.-M.~Richard, Phys. Rep. \textbf{368}, 119 (2002),
 \textit{Antinucleon-nucleon interaction at low energy: scattering and protonium}.

\bibitem{Entem2006}
D. R. Entem and F. Fern\'andez, 
Phys. Rev. C\textbf{73}, 045214 (2006),
\textit{The $N\bar N$ interaction in a constituent quark model: Baryonium and protonium level shifts}.

\bibitem{Trueman61}
T. L. Trueman,
Nucl. Phys. \textbf{26}, 57 (1961),
\textit{Energy level shifts in atomic states of strongly-interacting particles}.

\bibitem{Lambert70}
E. Lambert,
Helv. Phys. Acta \textbf{43}, 713 (1970),
\textit{Investigation of pionic atoms by means of scattering lengths}.

\bibitem{Carbonell92}
J. Carbonnell, J.-M. Richard, and S. Wycech,
Z. Phys. \textbf{A343}, 325 (1992),
\textit{On the relation between protonium level shifts and nucleon-antinucleon amplitudes}.

\bibitem{Entem2000}
D. R. Entem, F. Fern\'andez, and A. Valcarce,
Phys. Rev. C\textbf{62}, 034002 (2000),
\textit{ Chiral quark model of the $NN$ system within a Lippmann-Schwinger resonating group method}.

\bibitem{Lacombe84}
M. Lacombe, B. Loiseau, B.~Moussallam, and R.~Vinh~Mau,
Phys. Rev. C\textbf{29}, 1800 (1984),
\textit{Nucleon-antinucleon resonance spectrum in a potential model}.

\bibitem{Dedonder2008} J.-P. Dedonder, B. El-Bennich, 
B.~Loiseau, and S.~Wycech, arXiv:0904.2163v1 [nucl-th], 
submitted to Phys. Rev. C,
\textit{ On the structure of the $X(1835)$ baryonium}.

\bibitem{RVM73}
R. Vinh Mau, J-M. Richard, B. Loiseau, M. Lacombe and W.N. Cottingham,
Phys. Lett.  \textbf{44B}, 1 (1973),
\textit{Nucleon-Nucleon interaction from pion-nucleon phase shift analysis. The NN peripheral partial waves}.

\bibitem{Ablikim2005}
 M. Ablikim \textsl{et al.}, BES Collaboration, Phys. Rev. Lett. \textbf{95}, 262001 (2005). 
 {\it{ Observation of a resonance X(1835) in $ J/\psi \to \gamma \pi^{+}\pi^{-}\eta'$.}}

\end{thebibliography}
\end{document}